# Power Control in Networks With Heterogeneous Users: A Quasi-Variational Inequality Approach

Ivan Stupia, *Member, IEEE*, Luca Sanguinetti, *Member, IEEE*, Giacomo Bacci, *Member, IEEE*, Luc Vandendorpe, *Fellow, IEEE*

*Abstract*—This work deals with the power allocation problem in a multipoint-to-multipoint network, which is heterogenous in the sense that each transmit and receiver pair can arbitrarily choose whether to selfishly maximize its own rate or energy efficiency. This is achieved by modeling the transmit and receiver pairs as rational players that engage in a non-cooperative game in which the utility function changes according to each player's nature. The underlying game is reformulated as a quasi variational inequality (QVI) problem using convex fractional program theory. The equivalence between the QVI and the non-cooperative game provides us with all the mathematical tools to study the uniqueness of its Nash equilibrium points and to derive novel algorithms that allow the network to converge to these points in an iterative manner, both with and without the need for a centralized processing. Numerical results are used to validate the proposed solutions in different operating conditions.

## I. Introduction

The vision of seamless and pervasive wireless communication systems has paved the way to an extraordinary proliferation of wireless network infrastructures and ubiquitous services [1]. In this challenging arena, we consider a multipoint-to-multipoint network in which each transmit and receiver pair (TRP) can arbitrarily choose whether to selfishly maximize its own spectral efficiency (SE) (in terms of maximum achievable rate), or its own energy efficiency (EE) (in terms of trading off achievable rate and energy consumption). An example in which this heterogeneous multitude of users interact with each other might be represented by small-cell networks, which are founded on the idea of multiple radio access technologies, architectures and transmission techniques coexisting in the same area to ensure the most efficient usage of the spectrum resource with the minimum waste of energy [2].

Despite its promises, the deployment of small-cell networks poses several technical challenges mainly because different small cells are likely to be connected via an unreliable backhaul infrastructure whose features may strongly vary from case to case, with variable characteristics of error rate, delay, and capacity. This calls for developing flexible and decentralized power allocation strategies relying on local channel state information, and requiring only a small exchange of information.

I. Stupia and L. Vandendorpe are with the Université Catholique de Louvain, B-1348 Louvain-la-Neuve, Belgium (e-mail: ivan.stupia@uclouvain.be, luc.vanderdorpe@uclouvain.be).

L. Sanguinetti is with the University of Pisa, Dipartimento di Ingegneria dell'Informazione, Pisa, Italy and also with the Large Systems and Networks Group (LANEAS), CentraleSupélec, Gif-sur-Yvette, France (e-mail:luca.sanguinetti@unipi.it).

G. Bacci is with the University of Pisa, Dipartimento di Ingegneria dell'Informazione, Pisa, Italy (e-mail: giacomo.bacci@iet.unipi.it).

### A. Related works

As is well-known, a suitable tool to study and design complex interactions among rational entities operating in a distributed manner is *game theory*. In recent years, there has been a growing interest in adopting game theory to model many communications and networking problems (a good survey of the results on this topic can be found in [3]). Among the early contributions in this area, it is worth mentioning [4]–[6], in which the rate maximization problem for autonomous digital subscriber lines is addressed following a competitive optimality criterion. Following the route of such early works, most existing literature in wireless communications is focused on developing power control techniques for the maximization of the individual SE while satisfying individual power constraints. Some examples in the area of non-cooperative game theory are represented by the distributed power control strategies proposed in [7] for multi-user multiple-input multiple-output systems and by those developed in [8] and [9] for interference relay channels, whereas a two-tier network is considered in [10]. In [11], [12], the authors propose a detailed analysis of the Nash equilibrium (NE) point for the rate maximization problem in parallel Gaussian multiple access channels in which mobile users autonomously take decisions on the resource usage and compete with each other to exploit the available resources. More recently, in [13] the authors rely on the variational inequality (VI) framework to model and analyze the competitive rate maximization problem. The analogy between NE problems and VIs is also exploited in [14] to design distributed power control algorithms for rate maximization under interference temperature constraints in a cognitive radio context.

All the aforementioned distributed power allocation strategies have the great advantage of avoiding the excessive information exchange to achieve signal coordination as well as involved computational processing [13]. On the other hand, users' aggressive attitude towards interference can lead to a large transmission power at the mobile stations, thereby fostering an inefficient use of batteries. All this makes such existing solutions not suited for the development of energy-efficient networks. This has motivated a great interest in studying and designing resource allocation schemes taking into account the cost of energy in the performance metrics. Towards this end, the concept of link capacity per unit cost originally proposed by Verdú in [15] has been widely adopted in many different contexts (e.g., see [16]–[19] and references therein). In [19], [20], the authors focus on the same scenario investigated in [11], [12] and study the NE problem for a group of players aiming at maximizing their own EE while satisfying power constraints or rate requirements. Although interesting, the theoretical framework developed in [19], [20] does not provide a systematic study of the relationship between the SE and the EE maximization problems. To the best of our knowledge, a unified framework accommodating both is still missing.

## B. Contributions

The aim of this paper is to fulfill the gap mentioned above using the quasi VI (QVI) framework, originally introduced by Bensoussan in [21] as a modeling tool to be used in different fields such as economics and biology (see also [22] and references therein). Unlike the traditional VI framework, that has been widely applied in wireless communications, the use of QVI theory to develop numerical algorithms is relatively recent [23]. In this work, we apply the QVI to study the solution properties of power allocation in a *heterogeneous* game defined as non-cooperative game in which the users can locally choose whether to pursue their own SE or their own EE. This allows us to overcome the main limitations of existing approaches, which fail to provide closed-form conditions on the uniqueness of the equilibrium points and on the convergence properties of iterative solutions. Towards this goal, a two-step approach is used. First, the EE maximization problem introduced in [19] is reformulated as a QVI using convex fractional programming theory [17]. The same approach is then exploited to reformulate the heterogeneous game as a QVI. This is *per-se* sufficient to elaborate some insights on the properties of the NE points and to provide us with all the mathematical tools to study the uniqueness of the NE points of the heterogeneous game, and the convergence properties of iterative algorithms. In particular, we first propose a *centralized* approach, which relies on an iterative method for solving QVIs whose convergence is guaranteed under mild assumptions. Then, we propose an alternative solution exploiting the equivalence between the QVI and a nonlinear complementary problem (NCP), which gives each TRP the possibility to reach the NE in a *distributed* manner without the need for any centralized processing. The developed solutions are then validated by means of extensive simulations.

## C. Organization

The remainder of this paper is organized as follows. In Section II, we introduce the signal model, some basic notations and the problem under investigation. We also review the available literature on both the SE-only and the EE-only maximization problems, with particular emphasis on the major limitations of classical approaches when studying the EE problem. Section III illustrates the mathematical steps to reformulate the EE-only game as a QVI. This approach is then used in Section III-B to formalize the heterogeneous game in which both SE and EE users coexist. The uniqueness conditions for the NE points of the heterogeneous game are studied in Section IV. The QVI framework is also used in Section V to derive and study the convergence properties of two different iterative algorithms for achieving the NE points. Numerical results are shown in Section VI whereas some concluding remarks and discussions are drawn in Section VII.

## D. Notation

The following notation is used throughout the paper. Matrices and vectors are denoted by boldface letters. The notation $[\mathbf{A}]_{i,k}$ is used to indicate the $(i,k)$th entry of the enclosed matrix $\mathbf{A}$, and $\mathbf{A} = \text{diag}\{a(n); \; n = 1, 2, \ldots, N\}$ denotes an $N \times N$ diagonal matrix with entries $a(n)$ along its main diagonal. $\prod_i \mathcal{X}_i$ denotes the Cartesian product of the sets $\mathcal{X}_i$, and $\succeq$ stands for the element-wise greater or equal relations. $\mathbf{1}_N$ and $\mathbf{0}_N$ are the $N$-dimensional all-one and all-zero vectors, respectively. $\|\mathbf{x}\|$ denotes the Euclidean norm of vector $\mathbf{x}$, and $\|\mathbf{A}\|$ denotes the induced norm of matrix $\mathbf{A}$. The notation $\mathbf{x} \perp \mathbf{y}$ stands for $\mathbf{x}^T \mathbf{y} = 0$ and $\nabla_{\mathbf{x}} f(\mathbf{x}, \mathbf{y})$ denotes the gradient vector of $f(\mathbf{x}, \mathbf{y})$ with respect to $\mathbf{x}$. In addition, $\lambda_{\max}(\mathbf{A})$ and $\lambda_{\min}(\mathbf{A})$ denote respectively the maximum and the minimum eigenvalue of matrix $\mathbf{A}$. Finally, $\mathbb{1}_k^{\mathcal{K}}$ denotes the indicator function of a set $\mathcal{K}$ and it is such that $\mathbb{1}_k^{\mathcal{K}} = 1$ if $k \in \mathcal{K}$ and zero otherwise while $[x]^+ = \max(0, x)$.

## II. NASH EQUILIBRIUM PROBLEMS

Consider a $K$-user $N$-parallel Gaussian interference channel, in which there are $K$ TRPs sharing $N$ parallel Gaussian subchannels, that might represent time or frequency bins. The channel transfer function over the $n$th subchannel between the transmitter $i$ and receiver $k$ is denoted by $H_{k,i}(n)$. The transmission strategy of each user $k$ is the power allocation vector $\mathbf{p}_k = [p_k(1), p_k(2), \ldots, p_k(N)]^T$ over the $N$ subchannels satisfying the following (local) transmit power constraints:

$$\mathcal{P}_k = \left\{ \mathbf{p}_k \in \mathbb{R}_+^N : \; h_k(\mathbf{p}_k) \leq 0 \right\} \tag{1}$$

where $h_k(\mathbf{p}_k)$ is an affine function of $\mathbf{p}_k$ given by

$$h_k(\mathbf{p}_k) = \mathbf{1}^T \mathbf{p}_k - P_k \tag{2}$$

with $P_k$ being the total power available at transmitter $k$. We assume that the $K$ TRPs do not cooperate with each other and that the multi-user interference is simply treated as additive colored noise at each receiver. Moreover, local perfect channel state information is available at both transmitter and receiver sides.

In the above circumstances, the maximum achievable rate on link $k$ for a specific power allocation profile $\mathbf{p} = [\mathbf{p}_1^T, \mathbf{p}_2^T, \ldots, \mathbf{p}_K^T]^T$ is given by

$$R_k(\mathbf{p}_k, \mathbf{p}_{-k}) = \sum_{n=1}^{N} \log \left( 1 + \frac{|H_{k,k}(n)|^2 p_k(n)}{\sigma_k^2(n) + \sum_{i \neq k} |H_{k,i}(n)|^2 p_i(n)} \right) \tag{3}$$

where $\sigma_k^2(n)$ is the noise variance over the $n$th subcarrier on link $k$ and $\mathbf{p}_{-k} = [\mathbf{p}_1^T, \mathbf{p}_2^T, \ldots, \mathbf{p}_{k-1}^T, \mathbf{p}_{k+1}^T, \ldots, \mathbf{p}_K^T]^T$ collects the power allocation vectors of all transmitters, except the $k$th one. Following [16]–[19], the EE $E_k(\mathbf{p}_k, \mathbf{p}_{-k})$ of the $k$th link can be computed as

$$E_k(\mathbf{p}_k, \mathbf{p}_{-k}) = \frac{R_k(\mathbf{p}_k, \mathbf{p}_{-k})}{\Psi_k + \mathbf{1}^T \mathbf{p}_k} \tag{4}$$

with $\Psi_k > 0$ being the radio frequency (RF) circuitry power consumed at transmitter $k$.

As mentioned earlier, one of the major objectives of this work is to provide a framework to study heterogeneous multi-user systems in which each user can choose whether to maximize its own SE or EE. To this end, we first recall some fundamental results for the NE problem in which each player (the TRP) maximizes its own SE (Section II-A). Then, the case of competitive players aiming at maximizing the EE of the link is introduced and its NE points are mathematically characterized (Section II-B), using a fractional programming approach. Finally, in Section II-C we combine the results from these two game formulations to properly formalize and study the heterogeneous problem sketched above.

## A. Rate maximization

The rate maximization problem refers to a system in which each TRPr aims at selfishly choosing the power allocation strategy that maximizes its own rate for a given set of other players' power profile. Mathematically, this amounts to jointly solving the following problems:

$$\max_{\mathbf{p}_k} \quad R_k(\mathbf{p}_k, \mathbf{p}_{-k}) \quad \forall k \tag{5}$$
$$\text{subject to} \quad \mathbf{p}_k \in \mathcal{P}_k.$$

As is known, the joint solution of (5) $\mathbf{p}^\star = [\mathbf{p}_1^\star, \ldots, \mathbf{p}_K^\star] = [\mathbf{p}_k^\star, \mathbf{p}_{-k}^\star]$, such that $\mathbf{p}_k^\star = \arg\max_{\mathbf{p}_k \in \mathcal{P}_k} R_k(\mathbf{p}_k, \mathbf{p}_{-k}^\star)$, corresponds to the NE of the non-cooperative game with complete information defined as $\mathcal{G}_R = \langle \mathcal{K}, \{\mathcal{P}_k\}, \{R_k\}\rangle$ in which: $\mathcal{K} = \{1, 2, \ldots, K\}$ is the set of players; $\mathcal{P}_k$ denotes the strategy set of player $k$, defined as in (1); and $R_k$ is player $k$'s payoff function that is the rate defined in (3).

**Proposition 1** ([11]). *The Nash equilibria $\mathbf{p}^\star$ of $\mathcal{G}_R$ are found to be the fixed points of the waterfilling mappings given by*

$$p_k^\star(n) = \left[\mathsf{wf}_k(\mathbf{p}_{-k}^\star, \mu_k^\star)\right]_n \tag{6}$$

*where*

$$\left[\mathsf{wf}_k(\mathbf{p}_{-k}^\star, \mu_k^\star)\right]_n = \left[\frac{1}{\mu_k^\star} - \frac{\sigma_k^2(n) + \sum_{i \neq k} |H_{k,i}(n)|^2 p_i^\star(n)}{|H_{k,k}(n)|^2}\right]^+ \tag{7}$$

*and $\mu_k^\star$ is the water level, chosen such that*

$$\mathbf{1}^T \mathbf{p}_k^\star = P_k. \tag{8}$$

In [13], the authors provide a convenient way to study the properties of $\mathbf{p}^\star$ by showing that the rate maximization NE problem is equivalent to the nonlinear VI $\mathsf{VI}(\mathcal{P}, \mathbf{F})$ in which $\mathbf{F}(\mathbf{p}) = \{\mathbf{F}_k(\mathbf{p})\}_{k=1}^K$, with

$$\mathbf{F}_k(\mathbf{p}) = -\nabla_{\mathbf{p}_k} R_k(\mathbf{p}_k, \mathbf{p}_{-k}) \tag{9}$$

$$= \left\{-\left(\xi_k(n) + \sum_{i=1}^K D_{k,i}(n) p_i(n)\right)^{-1}\right\}_{n=1}^N \tag{10}$$

with

$$\xi_k(n) = \frac{\sigma_k^2(n)}{|H_{k,k}(n)|^2} \quad \text{and} \quad D_{k,i}(n) = \frac{|H_{k,i}(n)|^2}{|H_{k,k}(n)|^2}. \tag{11}$$

Thanks to the equivalence between the NE problem and the $\mathsf{VI}(\mathcal{P}, \mathbf{F})$, the following result can be proved.

**Theorem 1** ([13]). *A power allocation profile $\mathbf{p}^\star \in \mathcal{P}$ is an NE of the rate maximization problem if and only if*

$$(\mathbf{p} - \mathbf{p}^\star)^T \mathbf{F}(\mathbf{p}^\star) \geq 0 \quad \forall \mathbf{p} \in \mathcal{P} \tag{12}$$

*with $\mathcal{P} = \prod_{k=1}^K \mathcal{P}_k$.*

Interestingly, [11] also shows that the NE in (6) can be interpreted as a set of Euclidean projections onto the facets of $K$ polytopes. To see how this comes about, let $\boldsymbol{\xi}_k = [\xi_k(1), \xi_k(2), \ldots, \xi_k(N)]^T$ and $\mathbf{D}_{k,i} = \mathrm{diag}\{D_{k,i}(n); n = 1, 2, \ldots, N\}$. Define also $\overline{\mathcal{P}} = \prod_{k=1}^K \overline{\mathcal{P}}_k$ where each $\overline{\mathcal{P}}_k$ is the simplex

$$\overline{\mathcal{P}}_k = \{\mathbf{p}_k \in \mathbb{R}_+^N : h_k(\mathbf{p}_k) = 0\} \tag{13}$$

obtained from (1) when the power constraint is satisfied with equality.

**Proposition 2** ([12]). *A power allocation profile $\mathbf{p}^\star$ is an NE of the rate maximization problem if and only if*

$$\mathbf{p}_k^\star = \Pi_{\overline{\mathcal{P}}_k}\left(-\boldsymbol{\xi}_k - \sum_{i \neq k} \mathbf{D}_{k,i} \mathbf{p}_i^\star\right) \quad \forall k \tag{14}$$

*where $\Pi_{\overline{\mathcal{P}}_k}(\mathbf{z})$ computes the vector in the simplex $\overline{\mathcal{P}}_k$ that is closest to $\mathbf{z}$ in the Euclidean norm.*

One of the major advantages of reformulating the NE problem as in (14) is that the sufficient conditions for the uniqueness of the NE of $\mathcal{G}_R$ can be derived by simply studying the contraction property of $\Pi_{\overline{\mathcal{P}}}(\mathbf{z}) = [\Pi_{\overline{\mathcal{P}}_1}(\mathbf{z}_1) \cdots \Pi_{\overline{\mathcal{P}}_K}(\mathbf{z}_k)]^T$, where $\mathbf{z}_k = -\boldsymbol{\xi}_k - \sum_{i \neq k} \mathbf{D}_{k,i} \mathbf{p}_i$, with respect to the vector $\mathbf{p}$. In addition, the analysis of the convergence of iterative waterfilling-inspired algorithms is greatly simplified (please refer to [12] for more details on this subject). As we shall see, similar results can be proved to hold true for the energy-efficient maximization problem.

## B. Energy-efficiency maximization

The EE maximization problem refers to a network in which each player $k$ aims at selfishly choosing a power vector $\mathbf{p}_k \in \mathcal{P}_k$ to maximize its own EE $E_k(\mathbf{p}_k, \mathbf{p}_{-k})$ for a given set of other players' powers $\mathbf{p}_{-k}$. The problem can be mathematically formulated as:

$$\max_{\mathbf{p}_k} \quad E_k(\mathbf{p}_k, \mathbf{p}_{-k}) = \frac{R_k(\mathbf{p}_k, \mathbf{p}_{-k})}{\Psi_k + \mathbf{1}^T \mathbf{p}_k} \quad \forall k \tag{15}$$
$$\text{subject to} \quad \mathbf{p}_k \in \mathcal{P}_k.$$

Analogously to what is introduced in Section II-A, the solution of (15) is the NE of the noncooperative game $\mathcal{G}_E = \langle \mathcal{K}, \{\mathcal{P}_k\}, \{E_k\}\rangle$. Since the utility functions $E_k(\mathbf{p}_k, \mathbf{p}_{-k})$, $\forall k \in \mathcal{K}$, are strictly quasiconcave and $\mathcal{P}_k$ is a convex set, the players' best response to the opponent strategies can be computed using different convex optimization tools. Although possible, this direct approach, originally pursued in [19], presents some disadvantages, mainly because it does not bring any insights into the structure of the equilibrium points. This makes it hard to provide closed-form conditions for the uniqueness of the NE points of $\mathcal{G}_E$ and to study the

**Algorithm 1: Dinkelbach method**

**Data.** Set $i = 0$ and $\nu_k^{(0)} = \bar{\nu}$, with a random $\bar{\nu} > 0$. Choose $\epsilon \ll 1$.

**Step 1.** Compute
$$\mathbf{z}_k^{(i)}(\nu_k^{(i)}) = \mathsf{wf}_k(\mathbf{p}_{-k}, \nu_k^{(i)}).$$

**Step 2.** Set
$$\nu_k^{(i+1)} = \frac{R_k\left(\mathbf{z}_k^{(i)}(\nu_k^{(i)}), \mathbf{p}_{-k}\right)}{\Psi_k + \mathbf{1}^T \mathbf{z}_k^{(i)}(\nu_k^{(i)})}.$$

**Step 3.** If the condition
$$\left| R_k\left(\mathbf{z}_k^{(i)}(\nu_k^{(i)}), \mathbf{p}_{-k}\right) - \nu_k^{(i+1)}\left(\Psi_k + \mathbf{1}^T \mathbf{z}_k^{(i)}(\nu_k^{(i)})\right) \right| < \epsilon$$
is satisfied, then return $\nu_k^\star = \nu_k^{(i+1)}$ and STOP; otherwise, go to Step 4.

**Step 4.** Set $i \leftarrow i + 1$; and go back to Step 1.

---

convergence properties of iterative algorithms based on best response dynamics (see also [20]).

An alternative route (e.g., followed in [17]) relies on observing that (15) belongs to the class of concave-convex fractional programs, since $R_k(\mathbf{p}_k, \mathbf{p}_{-k})$ is a concave function of $\mathbf{p}_k$ whereas $\Psi_k + \mathbf{1}^T \mathbf{p}_k$ is affine and positive. Interestingly, the solution of such problems can be computed through methods that rely on different convex reformulations or duality approaches (see [24] for more details on this subject). Although different in principle, all these methods are very closely related to each other since they all lead to the same optimality condition. Following the parameter-free convex fractional program approach (whose main steps are reported in Appendix A for the sake of completeness), it turns out that player $k$'s best response $\mathcal{B}_k(\mathbf{p}_{-k})$ to an opponents' vector $\mathbf{p}_{-k}$ takes the form:

$$\mathcal{B}_k(\mathbf{p}_{-k}) = \mathsf{wf}_k(\mathbf{p}_{-k}, \lambda_k^\star(\mathbf{p}_{-k})) \tag{16}$$

where $\lambda_k^\star(\mathbf{p}_{-k})$ must satisfy the condition

$$\mathbf{1}^T \mathcal{B}_k(\mathbf{p}_{-k}) = \min\left\{ P_k, \frac{1}{t_k^\star(\mathbf{p}_{-k})} - \Psi_k \right\}. \tag{17}$$

The value of $1/t_k^\star(\mathbf{p}_{-k})$ corresponds to the total (radiated and fixed) power consumption for any given $\mathbf{p}_{-k}$ when the constraint $h_k(\mathbf{p}_k) \leq 0$ in (15) is neglected, and can be computed as (see Appendix A)

$$t_k^\star(\mathbf{p}_{-k}) = \frac{1}{\Psi_k + \mathbf{1}^T \mathbf{z}_k(\nu_k^\star)} \tag{18}$$

where $\mathbf{z}_k(\nu_k^\star)$ is given by

$$\mathbf{z}_k(\nu_k^\star) = \mathsf{wf}_k(\mathbf{p}_{-k}, \nu_k^\star) \tag{19}$$

with $\nu_k^*$ being iteratively obtained via Algorithm 1 (originally proposed by Dinkelbach in [25]) and such that

$$R_k\left(\mathbf{z}_k(\nu_k^\star), \mathbf{p}_{-k}\right) - \nu_k^\star\left(\Psi_k + \mathbf{1}^T \mathbf{z}_k(\nu_k^\star)\right) = 0. \tag{20}$$

Using the above results, the following proposition easily follows from the observation that a point $\mathbf{p}^\star$ is an NE if and only if $\mathbf{p}^\star \in \mathcal{B}(\mathbf{p}^\star)$, with $\mathcal{B}(\mathbf{p}) = \prod_{k=1}^K \mathcal{B}_k(\mathbf{p}_{-k})$.

**Proposition 3.** *The Nash equilibria $\mathbf{p}^\star$ of $\mathcal{G}_E$ are obtained as the fixed-point solutions of the following waterfilling mapping:*

$$\mathbf{p}_k^\star = \mathcal{B}_k(\mathbf{p}_{-k}^\star) = \mathsf{wf}_k(\mathbf{p}_{-k}^\star, \lambda_k^\star(\mathbf{p}_{-k}^\star)) \tag{21}$$

*with $\lambda_k^\star(\mathbf{p}_{-k}^\star)$ being such that*

$$\mathbf{1}^T \mathbf{p}_k^\star = \min\left\{ P_k, \frac{1}{t_k^\star(\mathbf{p}_{-k}^\star)} - \Psi_k \right\}. \tag{22}$$

**Remark 1.** *Similarly to $\mathcal{G}_R$, the NE points of $\mathcal{G}_E$ are found to be the fixed points of a waterfilling mapping, with the only difference that the water level must be chosen so as to satisfy (22) rather than (8).*

**Remark 2.** *Since $t_k^\star(\mathbf{p}_{-k})$ plays a major role in all subsequent discussions, let us point out its physical meaning and properties. As mentioned before, $1/t_k^\star(\mathbf{p}_{-k}) \in [\Psi_k, \infty)$ represents the total power dissipation that is required to maximize the EE for any given $\mathbf{p}_{-k}$ when the constraint $h_k(\mathbf{p}_k) \leq 0$ is neglected. Mathematically, $t_k^\star(\mathbf{p}_{-k})$ is obtained as the $t_k$-part of the solution of the following optimization problem (see Appendix A):*

$$\max_{\{\mathbf{y}_k, t_k\} \in \mathbb{R}_+^{N+1}} \quad t_k R_k(\mathbf{y}_k/t_k, \mathbf{p}_{-k}) \tag{23}$$
$$\text{subject to} \quad t_k(\Psi_k + \mathbf{1}^T \mathbf{y}_k/t_k) = 1.$$

*where $\mathbf{p}_k = \mathbf{y}_k/t_k$. From the above problem, it turns out that $1/t_k^\star(\mathbf{p}_{-k}) - \Psi_k$ is the total radiated power that would be needed by player $k$ to maximize its own EE for a given $\mathbf{p}_{-k}$. The maximum power constraint (1) acts as an upper bound to the strategy $\mathbf{p}_k^\star$, as follows from (17).*

It is worth observing that computing the NE as in (21), although useful to characterize its structure, does not provide any particular advantage in deriving conditions for the uniqueness of the NE points of $\mathcal{G}_E$. Similarly, the analysis on the convergence properties of the resulting iterative solutions is still much open. For this reason, analogously to what was done in [13] for the rate maximization game using the VI approach, in Section III we will make use of the above results to reformulate $\mathcal{G}_E$ as a QVI to exploit the powerful tools provided by the QVI theory. Before delving into this, we briefly introduce the heterogeneous maximization problem that will be analyzed in Section III-B.

### C. Heterogeneous maximization

Consider now a heterogeneous scenario in which a set $\mathcal{K}_R$ of players follows a rate maximization strategy, while the remaining set $\mathcal{K}_E$ is interested in maximizing its own EE. Let us define $\mathcal{G} = \langle \mathcal{K}, \{\mathcal{P}_k\}, \{u_k\} \rangle$ the corresponding game in which: $\mathcal{K} = \mathcal{K}_R \cup \mathcal{K}_E$ is the set collecting both types of users; the strategy set $\mathcal{P}_k$ is defined as in (1); and $u_k(\mathbf{p}_k, \mathbf{p}_{-k})$ is the utility function, defined as

$$u_k(\mathbf{p}_k, \mathbf{p}_{-k}) = \begin{cases} R_k(\mathbf{p}_k, \mathbf{p}_{-k}) & \text{if } k \in \mathcal{K}_R \\ E_k(\mathbf{p}_k, \mathbf{p}_{-k}) & \text{if } k \in \mathcal{K}_E. \end{cases} \tag{24}$$

The problem to be solved for each player $k$ can thus be mathematically formalized as follows:

$$\max_{\mathbf{p}_k} \quad u_k(\mathbf{p}_k, \mathbf{p}_{-k}) \quad \forall k \tag{25}$$
$$\text{subject to} \quad \mathbf{p}_k \in \mathcal{P}_k$$

and the corresponding Nash equilibria are obtained as follows using the results of Propositions 1 and 3.

**Proposition 4.** *The Nash equilibria $\mathbf{p}^\star$ of $\mathcal{G}$ are obtained as the fixed-point solutions of the following waterfilling mapping:*

$$\mathbf{p}_k^\star = \mathsf{wf}_k(\mathbf{p}_{-k}^\star, \lambda_k^\star(\mathbf{p}_{-k}^\star)) \tag{26}$$

*with $\lambda_k^\star(\mathbf{p}_{-k}^\star)$ being such that the following equality holds true:*

$$\mathbf{1}^T \mathbf{p}_k^\star = \begin{cases} P_k & \text{if } k \in \mathcal{K}_R \\ \min\left\{P_k, \dfrac{1}{t_k^\star(\mathbf{p}_{-k}^\star)} - \Psi_k\right\} & \text{if } k \in \mathcal{K}_E. \end{cases} \tag{27}$$

Similarly to the energy-efficient maximization problem, characterizing the properties of the NE points of $\mathcal{G}$, such as uniqueness conditions, and developing iterative solutions to achieve these points in a distributed manner, are open problems that call for an alternative approach. As mentioned earlier, we address all these issues by reformulating $\mathcal{G}$ as a QVI.

## III. QVI Formulation of $\mathcal{G}_E$

Next, we make use of the results of Section II-B to show how $\mathcal{G}_E$ can be reformulated as a QVI. To this end, for any given $\mathbf{p}_{-k}$, let us introduce the function $g_k(\mathbf{p}_k, \mathbf{p}_{-k})$, defined as

$$g_k(\mathbf{p}_k, \mathbf{p}_{-k}) = \mathbf{1}^T \mathbf{p}_k - \left(\dfrac{1}{t_k^\star(\mathbf{p}_{-k})} - \Psi_k\right) \tag{28}$$

where $t_k^\star(\mathbf{p}_{-k})$ is computed through (18)-(20). Let us also denote $\mathcal{P}_{-k} = \prod_{i \neq k} \mathcal{P}_i$ and

$$\mathcal{Q}(\mathbf{p}) = \prod_{k=1}^{K} \mathcal{Q}_k(\mathbf{p}_{-k}) \tag{29}$$

where $\mathcal{Q}_k : \mathcal{P}_{-k} \to 2^{\mathcal{P}_k}$, $\forall k$ collects the set-valued functions given by

$$\mathcal{Q}_k(\mathbf{p}_{-k}) = \mathcal{P}_k \cap \{\mathbf{p}_k \in \mathbb{R}_+^N : g_k(\mathbf{p}_k, \mathbf{p}_{-k}) \leq 0\} \tag{30}$$

whereas $2^{\mathcal{P}_k}$ is the power set collecting all the possible subsets of $\mathcal{P}_k$ [26]. Then, the following result can be obtained.

**Theorem 2.** *A power allocation profile $\mathbf{p}^\star \in \mathcal{Q}(\mathbf{p}^\star)$ is an NE of the EE maximization problem given in (15) if and only if it solves the* QVI$(\mathcal{Q}, \mathbf{F})$, *where $\mathbf{F}(\mathbf{p}) = \{\mathbf{F}_k(\mathbf{p})\}_{k=1}^K$, and $\mathbf{F}_k(\mathbf{p})$ is the mapping defined in (9). Stated formally, $\mathbf{p}^\star$ is an NE of $\mathcal{G}_E$ if and only if it is such that*

$$(\mathbf{p} - \mathbf{p}^\star)^T \mathbf{F}(\mathbf{p}^\star) \geq 0 \quad \forall \mathbf{p} \in \mathcal{Q}(\mathbf{p}^\star). \tag{31}$$

*Proof:* The proof is given in Appendix B and relies on proving that the Karush-Kuhn-Tucker (KKT) conditions of QVI$(\mathcal{Q}, \mathbf{F})$ are satisfied if and only if there exists a vector $\mathbf{p}^\star$ and a suitable Lagrange multiplier $\boldsymbol{\lambda}^\star \in \mathbb{R}_+^K$ such that (21) and (22) hold true. ∎

**Remark 3.** *The results of Theorem 2 are reminiscent of those in [14] in which the authors make use of the VI framework to solve the rate maximization problem in a cognitive radio network under interference constraints. In that case, the individual strategy sets $\mathcal{Q}_k(\mathbf{p}_{-k})$, $\forall k \in \mathcal{K}$, are defined as*

$$\mathcal{Q}_k(\mathbf{p}_{-k}) = \mathcal{P}_k \cap \{\mathbf{p}_k \in \mathbb{R}_+^N : I(\mathbf{p}_k, \mathbf{p}_{-k}) \leq 0\} \tag{32}$$

*where $I(\mathbf{p}_k, \mathbf{p}_{-k})$ denotes the interference constraint. The latter turns out to be the same for any link $k$ and moreover it is shown to be convex with respect to the entire strategy profile $\mathbf{p}$. This property allows the authors in [14] to reformulate the rate maximization problem as a VI. A close inspection of (28) reveals that $g_k(\mathbf{p}_k, \mathbf{p}_{-k})$ depends on the index link $k$ and it is not jointly convex with respect to the power allocation profile $\mathbf{p}$. For this reason the VI approach adopted in [14] cannot be applied to the problem at hand while the QVI framework is shown to be of great help.*

### A. Energy-efficiency maximization as a projector

For the sake of completeness, we observe that (similarly to what has been done for the rate maximization problem in [11]) the NE of $\mathcal{G}_E$ can also be interpreted as the Euclidean projection of the vector $-\boldsymbol{\xi}_k - \sum_{i \neq k} \mathbf{D}_{k,i} \mathbf{p}_i$ onto the simplex

$$\overline{\mathcal{Q}}_k(\mathbf{p}_{-k}) = \{\mathbf{p}_k \in \mathbb{R}_+^N : \max\{h_k(\mathbf{p}_k), g_k(\mathbf{p}_k, \mathbf{p}_{-k})\} = 0\} \tag{33}$$

where $g_k(\mathbf{p}_k, \mathbf{p}_{-k}) = 0$ is an affine hyperplane in $\mathbf{p}_k$ for any given $\mathbf{p}_{-k}$. This is summarized in the following proposition.

**Proposition 5.** *A power allocation profile $\mathbf{p}^\star$ is an NE of $\mathcal{G}_E$ if*

$$\mathbf{p}_k^\star = \Pi_{\overline{\mathcal{Q}}_k(\mathbf{p}_{-k}^\star)}\left(-\boldsymbol{\xi}_k - \sum_{i \neq k} \mathbf{D}_{k,i} \mathbf{p}_i^\star\right). \tag{34}$$

*Proof:* The proof is given in Appendix C and relies on proving that the KKT conditions of (34) are satisfied by the solution to (15). ∎

Capitalizing on the interpretation of the NE point as a Euclidean projection, a graphical comparison between the spectral-efficient (i.e., rate-maximizing) and energy-efficient best responses is provided in Fig. 1 for the case $N = 2$. In particular, Fig. 1 shows the player $k$'s best response to two different strategies of the other players, namely, $\mathbf{p}_{-k}$ and $\mathbf{p}'_{-k}$. As seen, the player $k$'s best responses are obtained as the projections of the interference vectors $-\boldsymbol{\xi}_k - \sum_{i \neq k} \mathbf{D}_{k,i} \mathbf{p}_i$ and $-\boldsymbol{\xi}_k - \sum_{i \neq k} \mathbf{D}_{k,i} \mathbf{p}'_i$ onto the corresponding hyperplanes: $h_k(\mathbf{p}_k) = 0$ for rate maximization, and $g_k(\mathbf{p}_k, \mathbf{p}_{-k}) = 0$ for energy-efficient maximization.[1] In sharp contrast to $h_k(\mathbf{p}_k) = 0$, the hyperplane $g_k(\mathbf{p}_k, \mathbf{p}_{-k}) = 0$ (corresponding to the straight lines below the grey regions) depends on the other players' strategies through $1/t_k^\star(\mathbf{p}_{-k})$, which varies within the interval $[\Psi_k, \infty)$. It is worth observing that all admissible power allocation profiles lie into the grey zone between the two hyperplanes. Due to space limitations, we cannot provide more details on this interpretation of the best response. We just observe that it can be useful to get further insights into the energy-aware optimization problem and more specifically into the trade-off between SE and EE policies in a competitive environment. Moreover, it might turn useful in some cases to study the convergence properties of distributed solutions as happened for the rate maximization problem (please refer to [12] for more details on this subject). However, this is left for future work and not pursued further in the sequel.

---

[1] If the projection of the interference vector is outside the grey region, then all the power is allocated only over the subchannel with the highest gain.

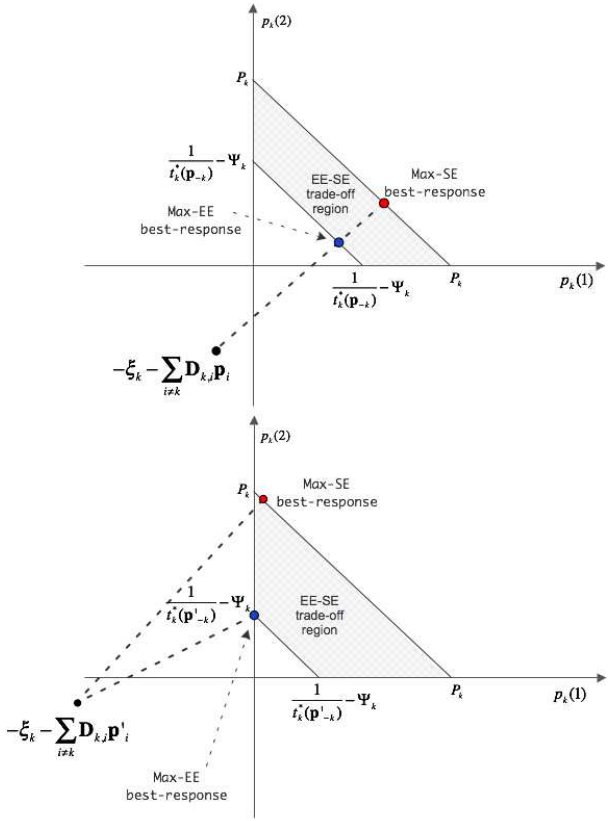

Fig. 1: Graphical illustration of SE and EE best responses.

### B. QVI Formulation of $\mathcal{G}$

Let us consider now the heterogeneous scenario. The following result can be proved.

**Proposition 6.** *Let $\mathcal{S}_k$ denote the set given by*

$$\mathcal{S}_k(\mathbf{p}_{-k}) = \begin{cases} \mathcal{P}_k & \text{if } k \in \mathcal{K}_R \\ \mathcal{Q}_k(\mathbf{p}_{-k}) & \text{if } k \in \mathcal{K}_E \end{cases} \quad (35)$$

*and let $\mathcal{S}: \mathcal{P} \to 2^{\mathcal{P}}$ be the set-valued function defined as the following Cartesian product:*

$$\mathcal{S}(\mathbf{p}) = \prod_{k \in \mathcal{K}} \mathcal{S}_k(\mathbf{p}_{-k}). \quad (36)$$

*Then, the NE problem in (25) is equivalent to* QVI$(\mathcal{S}, \mathbf{F})$, *which equals to finding a vector $\mathbf{p}^\star$ such that*

$$(\mathbf{p} - \mathbf{p}^\star)^T \mathbf{F}(\mathbf{p}^\star) \geq 0 \quad \forall \mathbf{p} \in \mathcal{S}(\mathbf{p}^\star) \quad (37)$$

*where $\mathbf{F}: \mathbb{R}^{NK} \to \mathbb{R}^{NK}$ is obtained as in (9).*

*Proof:* The proof follows the same steps as those used in Appendix B to prove Theorem 2. In particular, it relies on showing that the power allocation profile $\mathbf{p}_k$ satisfying the KKT conditions of the QVI is also a solution of the $k$th maximization problem in (25). ∎

### IV. ANALYSIS OF THE NASH EQUILIBRIA

The existence and uniqueness of the NE points of $\mathcal{G}$ are now studied.

**Proposition 7.** *The game $\mathcal{G}$ admits a nonempty set of NE points for any non-null maximum transmit power of users.*

*Proof:* The proof follows from observing that $\mathcal{G}$ satisfies the existence conditions in [27, Theorem 1.2]. In particular, the sets $\mathcal{P}_k$ are nonempty, convex and closed for any non-null maximum transmit power of users. Also, the payoff functions $R_k(\mathbf{p}_k, \mathbf{p}_{-k})$ and $E_k(\mathbf{p}_k, \mathbf{p}_{-k})$ are both quasi-concave with respect to $\mathbf{p}_k$. ∎

As far as the uniqueness of the NE is concerned, the following theorem provides a sufficient condition guaranteeing the uniqueness of the power allocation vector $\mathbf{p}^\star$ in (37) and thus in (25).

**Theorem 3.** *(Uniqueness conditions): Let $\mathbf{\Omega}(\cdot)$ be the mapping with elements given by*

$$[\mathbf{\Omega}(\mathbf{p})]_k = \begin{cases} P_k & \text{if } k \in \mathcal{K}_R \\ 1/t_k^\star(\mathbf{p}_{-k}) - \Psi_k & \text{if } k \in \mathcal{K}_E \end{cases} \quad (38)$$

*and let us define the matrices $\mathbf{A}$ and $\mathbf{B}$ whose elements are*

$$[\mathbf{A}]_{k,i} = \max_n \left\{ \frac{|H_{k,i}(n)|^2 |H_{k,k}(n)|^2}{\sigma_k^4(n)} \right\} \quad (39)$$

*and*

$$[\mathbf{B}]_{k,i} = \begin{cases} 1, & \text{if } i = k, \\ -\max_n \left\{ \frac{|H_{k,i}(n)|^2}{|H_{i,i}(n)|^2} \varsigma_{k,i}(n) \right\}, & \text{if } i \neq k, \end{cases} \quad (40)$$

*with $\varsigma_{k,i}(n)$ being defined as*

$$\varsigma_{k,i}(n) = \frac{\sigma_i^2(n) + \sum_\ell |H_{i,\ell}(n)|^2 P_\ell}{\sigma_k^2(n)}. \quad (41)$$

*The uniqueness of the NE in (25) is guaranteed under the following conditions.*

- *The matrix $\mathbf{B}$ is positive definite;*
- *There exists a nonnegative constant $\delta < 1/\Gamma$ such that*

$$\|\mathbf{\Omega}(\mathbf{p}) - \mathbf{\Omega}(\mathbf{p}')\| \leq \delta \|\mathbf{p} - \mathbf{p}'\| \quad \forall \mathbf{p}, \mathbf{p}' \in \mathcal{P} \quad (42)$$

*where*

$$\Gamma = \frac{\sqrt{\lambda_{\max}(\mathbf{A}^H \mathbf{A})}}{\lambda_{\min}(\widetilde{\mathbf{B}})} \max_k \max_n \left\{ (\tilde{\varsigma}_k(n))^2 \right\} \quad (43)$$

*is the so-called condition number of $\mathbf{F}$ with $\lambda_{\max}(\mathbf{A}^H \mathbf{A})$ and $\lambda_{\min}(\widetilde{\mathbf{B}})$ being the maximum eigenvalue of $\mathbf{A}^H \mathbf{A}$ and the minimum eigenvalue of the symmetric part of $\mathbf{B}$, respectively, and*

$$\tilde{\varsigma}_k(n) = \frac{\sigma_k^2(n)}{|H_{k,k}(n)|^2} + \sum_i \frac{|H_{k,i}(n)|^2}{|H_{k,k}(n)|^2} P_i. \quad (44)$$

**Algorithm 2:** Sequential penalty approach for solving QVI($\mathcal{S}, \mathbf{F}$).

**Data.** Choose an increasing sequence $\{\rho^{(i)}\}_{i=0}^{\infty}$ satisfying (48), and a sequence of vectors $\{\boldsymbol{\alpha}^{(i)}\}_{i=0}^{\infty}$. Set $j = 0$.

**Step 1.** Compute $\mathbf{p}^{(j)}$ as the solution of the penalized VI in (49).

**Step 2.** If a suitable termination criterion is satisfied, then return $\mathbf{p}^{\mathsf{SPA}} = \mathbf{p}^{(j)}$ and STOP, otherwise go to Step 3.

**Step 3.** Set $j \leftarrow j + 1$; and go back to Step 1.

*Proof:* The proof is given in Appendix D and relies on [28, Theorem 4.1]. Observe that, unlike [14], the positive definiteness of $\mathbf{B}$ is not sufficient to guarantee the uniqueness of the NE point (see discussion below). ∎

**Remark 4.** *Observe that when $\mathcal{K}_E = \emptyset$ and thus $\mathcal{K} = \mathcal{K}_R$ then (42) is always satisfied since $\|\boldsymbol{\Omega}(\mathbf{p}) - \boldsymbol{\Omega}(\mathbf{p}')\| = 0$. Therefore, the uniqueness conditions only require the matrix $\mathbf{B}$ to be positive definite. Not surprisingly, this result coincides with that in [14], wherein the authors show that the positive definiteness of $\mathbf{B}$ is a sufficient condition to claim the uniqueness of the NE for the rate maximization game. As rigorously discussed in [14], $\mathbf{B}$ is positive definite if for any $k$ one (or both) of the following conditions is fulfilled:*

$$\frac{1}{w_k}\sum_{i\neq k} w_i \max_n \left\{ \frac{|H_{k,i}(n)|^2}{|H_{i,i}(n)|^2} \varsigma_{k,i}(n) \right\} < 1 \quad (45)$$

$$\frac{1}{w_i}\sum_{k\neq i} w_k \max_n \left\{ \frac{|H_{k,i}(n)|^2}{|H_{i,i}(n)|^2} \varsigma_{k,i}(n) \right\} < 1 \quad (46)$$

*where $\mathbf{w} = [w_1, w_2, \ldots, w_K]^T$ is some positive vector. The above inequalities say that the positive definiteness of $\mathbf{B}$ is ensured if the received and/or generated multi-user interference is relatively low [14]. On the other hand, if $\mathcal{K}_E \neq \emptyset$ then*

$$\|\boldsymbol{\Omega}(\mathbf{p}) - \boldsymbol{\Omega}(\mathbf{p}')\|^2 = \sum_{k\in\mathcal{K}_E} ([\boldsymbol{\Omega}(\mathbf{p})]_k - [\boldsymbol{\Omega}(\mathbf{p}')]_k)^2 \quad (47)$$

*where $[\boldsymbol{\Omega}(\mathbf{p})]_k$ is the total power consumption on link $k$ given by (38). Therefore, the positive definiteness of $\mathbf{B}$ alone is no longer sufficient since (42) is not always satisfied. A close inspection of (42)-(43) leads to the following interpretation of this additional condition: The game $\mathcal{G}$ has a unique NE if at a variation of the opponent players' strategy corresponds a relatively small variation of the total power consumption in (47). This comes from the definition of the condition number of $\mathbf{F}$, which is formally defined as the value of the asymptotic worst-case relative change in the output for a relative change in the input. Observe that $\Gamma$ takes a finite value since $\mathbf{F}$ is a smooth function.*

## V. ITERATIVE ALGORITHMS TO SOLVE $\mathcal{G}$

In what follows, we show how to exploit the above theoretical framework to compute the NE of $\mathcal{G}$. Towards this goal, two different solutions are proposed. The first one exploits the equivalence between $\mathcal{G}$ and QVI($\mathcal{S}, \mathbf{F}$), and show how the solution of $\mathcal{G}$ can be computed by resorting to an iterative method for solving QVIs. As we shall see, this results into an iterative procedure that, in principle, requires a centralized implementation. The second approach relies on showing that QVI($\mathcal{S}, \mathbf{F}$) is equivalent to a nonlinear complementary problem (NCP), which gives each player the possibility to reach the NE of $\mathcal{G}$ in a distributed manner without the need for any centralized processing.

### A. A sequential penalty approach

The solution to $\mathcal{G}$ is next numerically obtained through the iterative procedure for solving QVIs known as sequential penalty approach (SPA) (e.g., see [23]). The latter is inspired by the augmented Lagrangian approach for nonlinear programming and its key idea is to solve the QVI by iteratively solving a properly defined sequence of penalized VIs on the set $\mathcal{P}$.

Specifically, let $j$ be the iteration index and let $\{\rho^{(j)}\} \subset \mathbb{R}_+$ be a sequence of given positive scalars satisfying $\rho^{(j)} < \rho^{(j+1)}$ and tending to $\infty$, i.e.,

$$\lim_{j\to\infty} \rho^{(j)} \to \infty. \quad (48)$$

Let also $\{\boldsymbol{\alpha}^{(j)}\} \subset \mathbb{R}^K$ be a sequence of some arbitrary vectors, and denote by $\mathbf{p}^{(j)}$ the solution of the following penalized VI$^{(j)}(\mathcal{P}, \mathbf{F} + \nabla \mathbf{C})$ on the set $\mathcal{P}$:

$$(\mathbf{p} - \mathbf{p}^{(j)})^T \left( \mathbf{F}(\mathbf{p}^{(j)}) + \nabla_{\mathbf{p}^{(j)}} \mathbf{C}(\mathbf{p}^{(j)}) \right) \geq 0 \quad \forall \mathbf{p} \in \mathcal{P} \quad (49)$$

where $\nabla_{\mathbf{p}^{(j)}} \mathbf{C}(\mathbf{p}^{(j)}) = \{\nabla_{\mathbf{p}_k^{(j)}} C_k(\mathbf{p}^{(j)})\}_{k=1}^{K}$ is the penalty mapping at the $j$th iteration, with

$$C_k(\mathbf{p}^{(j)}) = \frac{\mathbb{1}_k^{\mathcal{K}_E}}{2\rho^{(j)}} \left( \left[ \alpha_k^{(j)} + \rho^{(j)} g_k\left(\mathbf{p}_k^{(j)}, \mathbf{p}_{-k}^{(j)}\right) \right]^+ \right)^2. \quad (50)$$

The resulting iterative procedure is reported in Algorithm 2 and its convergence properties are stated in the following proposition.

**Proposition 8.** *Let $\mathbf{p}^{(j)}$ being the solution of (49) at iteration $j$. If $\{\boldsymbol{\alpha}^{(j)}\}$ is a bounded sequence of vectors, i.e.,*

$$\max_k \left|\alpha_k^{(j)}\right| < \bar{\alpha} \quad \forall j \quad (51)$$

*with $\bar{\alpha}$ some non-negative real-valued number, then*

$$\mathbf{p}^{\mathsf{SPA}} = \lim_{j\to\infty} \mathbf{p}^{(j)} \quad (52)$$

*is a solution of QVI($\mathcal{S}, \mathbf{F}$) and thus of $\mathcal{G}$.*

*Proof:* The proof is given in Appendix E and exploits the more general result provided by [23, Theorem 3] which, in turn, requires to prove that $g_k(\mathbf{p}_k, \mathbf{p}_{-k})$ is a continuous function with respect to the entire power vector $\mathbf{p}$. ∎

As seen, each iteration of Algorithm 2 requires only to solve a penalized VI. Next, we briefly illustrate how this can be done using standard optimization techniques. Let us introduce a slack variable $\lambda_k^{(j)}$ such that $\lambda_k^{(j)} = 0$ if $k \in \mathcal{K}_R$ and

$$0 \leq \lambda_k^{(j)} \perp (\lambda_k^{(j)} - \alpha_k^{(j)} - \rho^{(j)} g_k(\mathbf{p}_k^{(j)}, \mathbf{p}_{-k}^{(j)})) \geq 0 \quad (53)$$

**Algorithm 3: Distributed algorithm for solving $\mathsf{QVI}(\mathcal{S}, \mathbf{F})$**

**Data.** Set $j = 0$, $\boldsymbol{\gamma}^{(0)} = \mathbf{0}_K$ and $\boldsymbol{\Phi}(\boldsymbol{\gamma}^{(0)}) = \mathbf{0}_K$. Choose $\epsilon \ll 1$.

**Step 1.** Use $\boldsymbol{\gamma}^{(j)}$ to compute $\mathbf{p}^{\mathsf{VI}}(\boldsymbol{\gamma}^{(j)})$ solving

$$(\mathbf{p} - \mathbf{p}^{\mathsf{VI}}(\boldsymbol{\gamma}^{(j)}))^T \left( \mathbf{F}(\mathbf{p}^{\mathsf{VI}}(\boldsymbol{\gamma}^{(j)})) + \boldsymbol{\gamma}^{(j)} \right) \geq 0 \quad \forall \mathbf{p} \in \mathcal{P}$$

using the IWFP (e.g., via Algorithm 4).

**Step 2.** Use $\mathbf{p}^{\mathsf{VI}}(\boldsymbol{\gamma}^{(j)})$ to compute $t_k^\star(\mathbf{p}_{-k}^{\mathsf{VI}}(\boldsymbol{\gamma}^{(j)}))$ for any $k \in \mathcal{K}_E$ via the Dinkelbach method (e.g., via Algorithm 1).

**Step 3.** Use $\mathbf{p}_k^{\mathsf{VI}}(\boldsymbol{\gamma}^{(j)})$ and $t_k^\star(\mathbf{p}_{-k}^{\mathsf{VI}}(\boldsymbol{\gamma}^{(j)}))$ to set

$$\left[\boldsymbol{\Phi}(\boldsymbol{\gamma}^{(j)})\right]_k = \frac{1}{t_k^\star(\mathbf{p}_{-k}^{\mathsf{VI}}(\boldsymbol{\gamma}^{(j)}))} - \Psi_k - \mathbf{1}^T \mathbf{p}_k^{\mathsf{VI}}(\boldsymbol{\gamma}^{(j)})$$

for any $k \in \mathcal{K}_E$.

**Step 4.** If $\max_k \left| \gamma_k^{(j)} \left[\boldsymbol{\Phi}(\boldsymbol{\gamma}^{(j)})\right]_k \right| \leq \epsilon$, then return $\mathbf{p}^\star = \mathbf{p}^{\mathsf{VI}}(\boldsymbol{\gamma}^{(j)})$ and STOP, otherwise go to Step 5.

**Step 5.** Choose $\tau^{(j)} > 0$. Set

$$\gamma_k^{(j+1)} = \left[ \gamma_k^{(j)} - \tau^{(j)} \left[\boldsymbol{\Phi}(\boldsymbol{\gamma}^{(j)})\right]_k \right]^+$$

for any $k \in \mathcal{K}_E$.

**Step 6.** Set $j \leftarrow j + 1$; and go back to Step 1.

---

if $k \in \mathcal{K}_E$. From the KKT conditions of (49) and observing that the squared max function (50) is once continuously differentiable with

$$\left[ \nabla_{\mathbf{p}_k^{(j)}} C_k(\mathbf{p}^{(j)}) \right]_n = \left[ \alpha_k^{(j)} + \rho^{(j)} g_k\left(\mathbf{p}_k^{(j)}, \mathbf{p}_{-k}^{(j)}\right) \right]^+ \quad (54)$$

it follows that $\mathbf{p}^{(j)}$ is the solution of $\mathsf{VI}^{(j)}(\mathcal{P}, \mathbf{F} + \nabla \mathbf{C})$ if and only if there exist some vectors $\boldsymbol{\mu}^{(j)} \in \mathbb{R}_+^K$ such that the following stationarity condition is satisfied for any $k$:

$$\mathbf{F}_k(\mathbf{p}_k^{(j)}, \mathbf{p}_{-k}^{(j)}) + \lambda_k^{(j)} \mathbf{1}_N + \mu_k^{(j)} \mathbf{1}_N = \mathbf{0}_N \quad (55)$$

with $0 \leq \mu_k^{(j)} \perp h_k(\mathbf{p}_k^{(j)}) \leq 0$ and $\lambda_k^{(j)}$ defined above. Moreover, for any given $\boldsymbol{\lambda}^{(j)} = [\lambda_1^{(j)}, \lambda_2^{(j)}, \ldots, \lambda_K^{(j)}]^T$ and $\boldsymbol{\mu}^{(j)}$, the vector $\mathbf{p}^{(j)}$ solving (55) can be easily found in closed form as a function of $\boldsymbol{\lambda}^{(j)}$ and $\boldsymbol{\mu}^{(j)}$. Therefore, the solution of the $j$th iteration can be obtained solving a constrained system of equations in $\boldsymbol{\lambda}^{(j)}$ and $\boldsymbol{\mu}^{(j)}$. This can be done using standard methods for nonsmooth continuous equations [29, Chapter 8].

In general, the implementation of Algorithm 2 requires a centralized unit or an excessive exchange of information among the users. Although possible, this is clearly not suited for those applications (such as small-cell networks) wherein the exchange of information among the users is unreliable or even impossible. For this reason, in what follows we propose an alternative solution, which operates in a distributed manner and requires only local information.

### B. An NCP-based approach

Assume that $\boldsymbol{\gamma} \in \mathbb{R}_+^K$ is a given vector and denote $\mathbf{p}^{\mathsf{VI}}(\boldsymbol{\gamma})$ as the solution of the following penalized VI:

$$(\mathbf{p} - \mathbf{p}^{\mathsf{VI}}(\boldsymbol{\gamma}))^T \left( \mathbf{F}(\mathbf{p}^{\mathsf{VI}}(\boldsymbol{\gamma})) + \boldsymbol{\gamma} \right) \geq 0 \quad \forall \mathbf{p} \in \mathcal{P}. \quad (56)$$

---

**Algorithm 4: Sequential IWFP for a given $\boldsymbol{\gamma}^{(j)} \succeq \mathbf{0}$**

**Data.** Set $m = 0$ and choose any $\boldsymbol{\xi}_k^{(0)} \in \mathcal{P}_k$ for $k = 1, 2, \ldots, K$.

**Step 1.** Sequentially for $k = 1, 2, \ldots, K$ compute

$$\boldsymbol{\xi}_k^{(m)} = \mathsf{wf}_k \left( \boldsymbol{\xi}_{-k}^{(m)}, \chi_k^{(m)} + \gamma_k^{(j)} \right)$$

where $\mathsf{wf}_k$ is defined as in (7), and $\chi_k^{(m)}$ is chosen to satisfy the power constraint $\mathbf{1}_K^T \boldsymbol{\xi}_k^{(m)} = P_k$ if

$$\mathbf{1}_K^T \mathsf{wf}_k \left( \boldsymbol{\xi}_{-k}^{(m)}, \gamma_k^{(j)} \right) \geq P_k$$

and $\chi_k^{(m)} = 0$ otherwise.

**Step 2.** If a suitable termination criterion is satisfied, then $\mathbf{p}^{\mathsf{VI}}(\boldsymbol{\gamma}^{(j)}) = \boldsymbol{\xi}^{(m)}$ and STOP, otherwise go to Step 3.

**Step 3.** Set $m \leftarrow m + 1$; and go back to Step 1.

---

Let us also define the mapping $\boldsymbol{\Phi} : \mathbb{R}_+^K \to \mathbb{R}^K$ as

$$[\boldsymbol{\Phi}(\boldsymbol{\gamma})]_k = \begin{cases} 0 & \text{if } k \in \mathcal{K}_R \\ \frac{1}{t_k^\star(\mathbf{p}_{-k}^{\mathsf{VI}}(\boldsymbol{\gamma}))} - \Psi_k - \mathbf{1}^T \mathbf{p}_k^{\mathsf{VI}}(\boldsymbol{\gamma}) & \text{if } k \in \mathcal{K}_E \end{cases} \quad (57)$$

with $t_k^\star(\mathbf{p}_{-k}^{\mathsf{VI}}(\boldsymbol{\gamma}))$ being obtained via (18)-(20) after replacing $\mathbf{p}_{-k}$ with $\mathbf{p}_{-k}^{\mathsf{VI}}(\boldsymbol{\gamma})$. Then, the following result can be proven.

**Proposition 9.** *A power allocation profile $\mathbf{p}^\star$ is an NE of $\mathcal{G}$ if it solves (56) and $\boldsymbol{\gamma}$ is solution of the $\mathsf{NCP}(\boldsymbol{\Phi})$ given by:*

$$\mathsf{NCP}(\boldsymbol{\Phi}): \quad \mathbf{0} \leq \boldsymbol{\gamma} \perp \boldsymbol{\Phi}(\boldsymbol{\gamma}) \geq \mathbf{0}. \quad (58)$$

*Proof:* The proof is provided in Appendix F and relies on showing that the solution of $\mathsf{NCP}(\boldsymbol{\Phi})$ along with the solution of the penalized VI satisfies the KKT of $\mathsf{QVI}(\mathcal{S}, \mathbf{F})$, and thus $\mathbf{p}^\star$ is also an NE point of $\mathcal{G}$. ∎

Taking advantage of the above proposition, we develop the iterative scheme illustrated in Algorithm 3, which operates through a 2-layer procedure. More in details, at the $j$th iteration the inner layer represented by Step 1 makes use of $\boldsymbol{\gamma}^{(j)}$ to compute the solution of (56). This is achieved using the iterative waterfilling with pricing (IWFP) procedure proposed in [14] and reported in Algorithm 4 for completeness. The outer layer makes use of $\mathbf{p}_k^{\mathsf{VI}}(\boldsymbol{\gamma}^{(j)})$ (provided by Step 1) to update $\boldsymbol{\gamma}^{(j)}$ through Steps 2 – 5. The convergence properties of Algorithm 3 are stated in the following proposition, in which $\{\tau^{(j)}\}$ represent the scalar parameters used in Step 5.

**Proposition 10.** *Assume that:*

- *The matrix $\mathbf{B}$ defined in (40) is positive definite;*
- *The mapping $\boldsymbol{\Phi}(\boldsymbol{\gamma})$ is a co-coercive function of $\boldsymbol{\gamma}$ with constant $\kappa > 0$, i.e.,*

$$(\boldsymbol{\gamma}_1 - \boldsymbol{\gamma}_2)^T (\boldsymbol{\Phi}(\boldsymbol{\gamma}_1) - \boldsymbol{\Phi}(\boldsymbol{\gamma}_2)) \geq \kappa \|\boldsymbol{\Phi}(\boldsymbol{\gamma}_1) - \boldsymbol{\Phi}(\boldsymbol{\gamma}_2)\|^2 \quad (59)$$

*for any $\boldsymbol{\gamma}_1, \boldsymbol{\gamma}_2 \in \mathbb{R}_+^K$.*

*If the scalars $\tau^{(j)}$ are chosen such that*

$$0 \leq \inf_j \tau^{(j)} \leq \sup_j \tau^{(j)} \leq 2\kappa \quad (60)$$

then the sequence of vectors $\{\mathbf{p}^{\text{VI}}(\boldsymbol{\gamma}^{(j)})\}$ generated by Algorithm 3 converges to the NE of $\mathcal{G}$.

*Proof:* The convergence of the inner layer to the solution of (56) follows from the results in [14], in which the authors prove that if the mapping $\mathbf{F}$ is strongly monotone then $\mathbf{p}^{\text{VI}}(\boldsymbol{\gamma})$ can be computed in a distributed manner for any given $\boldsymbol{\gamma} \succeq \mathbf{0}$ through IWFP. As pointed out in Appendix D, the condition on the strong monotonicity of $\mathbf{F}$ holds true whenever the matrix $\mathbf{B}$ in (40) is positive definite. On the other hand, the convergence of the outer layer can be proved by simply observing that it is equivalent to the projection method with variable steps described in [29, Algorithm 12.1.4]. Therefore, the convergence proof follows from [29, Theorem 12.1.8]. ∎

Unlike Algorithm 2, Algorithm 3 enables the computation of the NE points of $\mathcal{G}$ in a distributed manner without the need for any centralized processing. To see how this comes about, observe that the evaluation of $\mathbf{p}_k^{\text{VI}}(\boldsymbol{\gamma}^{(j)})$ through IWFP in the inner layer only requires knowledge of the local measure of the overall interference plus noise. This information can easily be estimated at each transmitter during its own reception phase. The same information is needed in the outer layer by each player $k$ for updating the value of $t_k^\star(\mathbf{p}_{-k}^{\text{VI}}(\boldsymbol{\gamma}^{(j)}))$ in Step 2. Once $t_k^\star(\mathbf{p}_{-k}^{\text{VI}}(\boldsymbol{\gamma}^{(j)}))$ is computed, then $[\boldsymbol{\Phi}(\boldsymbol{\gamma}^{(j)})]_k$ is evaluated in Step 3 and later used in Step 5 to update $\gamma_k^{(j+1)}$.

**Remark 5** (On the design of $\{\tau^{(j)}\}$ in Algorithm 3). *From (60), it follows that a judicious design of $\{\tau^{(j)}\}$ would require to compute $\kappa$ in closed form as a function of the system parameters. Unfortunately, this is a challenging task, which is still much open. In Appendix G, we make use of some heuristic arguments to conjecture that if the uniqueness conditions of Theorem 3 are satisfied, then $\boldsymbol{\Phi}(\boldsymbol{\gamma})$ is a co-coercive function of $\boldsymbol{\gamma}$ with constant*

$$\kappa = \frac{\beta}{1 + \Gamma^{-2}} \quad (61)$$

*where*

$$\beta = \frac{\lambda_{\min}(\widetilde{\mathbf{B}})}{\max_k \max_n \{(\tilde{s}_k(n))^2\}} \quad (62)$$

*is the strong monotonicity constant of $\mathbf{F}$ while $\Gamma$ is its condition number (see also Appendix D). A formal proof of (61) is beyond the scope of this work, and it is currently under investigation. Herewith, we observe that, although building upon heuristics, this condition has been validated by means of the extensive simulations shown in the next section. In particular, it turns out that the proposed distributed algorithm converges with probability one whenever the uniqueness conditions depicted in Theorem 3 are satisfied.*

## VI. NUMERICAL RESULTS

Numerical results are now provided to assess the performance of the proposed solutions when applied to a heterogeneous network. In particular, we consider a scenario with $K = 8$ players in which four of them aim at maximizing the EE and the other four are focused on the SE maximization. The system parameters are as follows: *i)* the interference channel is composed of $N = 16$ subchannels; *ii)* the channel coefficients $H_{k,i}(n)$ are assumed to be $\mathcal{CN}(0,1)$ $\forall k, i, n$; *iii)* the average signal-to-noise ratio (SNR) on the generic subchannel $n$ over link $k$ is defined as $\text{SNR}_k(n) = \text{E}\{|H_{k,k}(n)|^2\}/\sigma_k^2$ and it is set to 0 dB $\forall k, n$; *iv)* the maximum normalized power is fixed to $P_k = N$ for any $k$; *v)* the static power consumption is assumed to be $\Psi_k = 1$ for all $k$; *vi)* the starting point of the distributed algorithms is the uniform power allocation strategy, i.e., $\mathbf{p}_k^{(0)} = \mathbf{1}$; *vii)* the tolerance parameter of the Dinkelbach's algorithm is set to $\epsilon = 10^{-6}$, *viii)* the sequence $\{\tau^{(j)}\}$ is chosen such that the upper bound in (60) is met with equality with $\kappa$ given by (61).

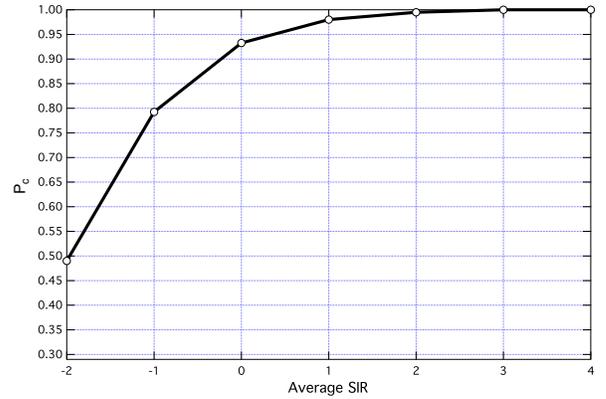

Fig. 2: Convergence probability of Algorithm 3.

Fig. 2 shows the empirical probability that Algorithm 3 converges to the solution of the QVI obtained via Algorithm 2. The convergence probability $P_c$ is plotted as a function of the average signal to interference ratio (SIR) on the generic subchannel $n$ over link $k$ defined as $\text{SIR}_k(n) = \text{E}\{|H_{k,k}(n)|^2\}/(\sum_{i \neq k} \text{E}\{|H_{k,i}(n)|^2\})$. As it is seen, in the low SIR regime Algorithm 3 does not always converge to the solution of the QVI and thus to the NE point of $\mathcal{G}$. This is because the co-coercivity of $\boldsymbol{\Phi}$ is not guaranteed for small values of SIRs. On the other hand, the convergence probability is numerically close to one for moderate-to-high values of SIR.

Figs. 3-4 show respectively the EE and the SE dynamics during the time interval needed by Algorithm 3 to converge. As can seen, a stable power allocation strategy is achieved while enhancing the EE of the energy efficient users up to the 113% with respect to static uniform power allocation at the price of a consistent information rate loss. The equilibrium is also achieved for the SE users with a final SE more than doubled with respect the initial uniform power allocation.

## VII. CONCLUSION

In this work, we have studied a non-cooperative game modelling the power allocation problem that arises in a heterogenous multipoint-to-multipoint network wherein each TRP can arbitrarily choose whether to selfishly maximize its own SE or EE. To overcome the main limitations of existing methodologies, we have reformulated the underlying game as a QVI problem and then we have exploited the powerful tools of the QVI theory: *i)* to study the uniqueness of the NE points; and *ii)* to derive novel algorithms to converge to the NE points in an iterative manner both with and without the need for a centralized processing. Numerical results have been used to validate the performance of the proposed solutions in a heterogeneous network.

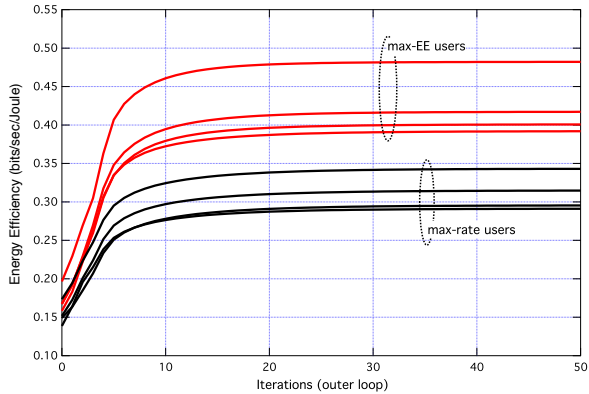

Fig. 3: Energy efficiency dynamics when $\text{SIR}_k(n) = 3$ dB.

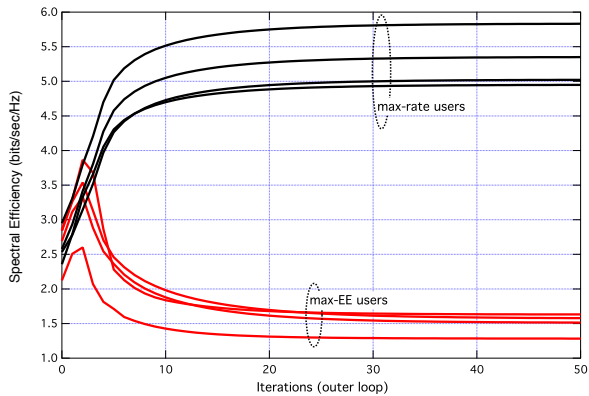

Fig. 4: Spectral efficiency dynamics when $\text{SIR}_k(n) = 3$ dB.

This work must be considered as a first attempt in using the QVI theory for dealing with EE in a competitive environment. We do believe that the developed framework will be of great help to deal with several interesting extensions (as the VI theory was useful to better study the rate maximization problem). For example, the above results might be in principle extended to any $t_k^\star(\mathbf{p}_{-k})$ that is a continuously differentiable function of the set of other players' powers $\mathbf{p}_{-k}$. This means that any additional constraint in (15) or (25) could be easily handled by $\text{QVI}(\mathcal{S}, \mathbf{F})$ whenever it can be incorporated in $t_k^\star(\mathbf{p}_{-k})$. This might be the case of minimum data rate requirement or maximum allowed interference levels.

Another interesting result that builds upon the developed framework is as follows. Exploiting the well-established relationship between QVIs and generalized NE problems (e.g., see [23], [30]), the heterogeneous game $\mathcal{G}$ turns out to be equivalent to $\mathcal{G}' = \langle \mathcal{K}, \{\mathcal{S}_k\}, \{R_k\} \rangle$ in which the problem to be solved for each player $k$ takes the form:

$$\max_{\mathbf{p}_k} \quad R_k(\mathbf{p}_k, \mathbf{p}_{-k}) \qquad (63)$$
$$\text{subject to} \quad \mathbf{p}_k \in \mathcal{S}_k(\mathbf{p}_{-k}).$$

This means that a heterogeneous game $\mathcal{G}$ in which each player can choose whether to maximize its own SE or EE is basically equivalent to a game $\mathcal{G}'$ in which the utility to maximize is always the rate but the strategy set of each player depends on the opponents' strategies $\mathbf{p}_{-k}$ through $\mathcal{S}_k(\mathbf{p}_{-k})$. This result may be useful to get valuable insights into the tradeoff between SE and EE in competitive environments.

## APPENDIX A
## PROBLEM (15) AS A FRACTIONAL PROGRAM

In this appendix, the solution of (15) is computed by means of the parameter-free convex fractional program approach (e.g., see [17]). To this end, we set

$$\mathbf{y}_k = \frac{\mathbf{p}_k}{\Psi_k + \mathbf{1}^T \mathbf{p}_k} \qquad (64)$$

and define $t_k$ as the inverse of the total power dissipation

$$t_k = \frac{1}{\Psi_k + \mathbf{1}^T \mathbf{p}_k}. \qquad (65)$$

Then, the $k$th maximization problem in (15) can be rewritten in the following parameter-free form:

$$\max_{\{\mathbf{y}_k, t_k\} \in \mathbb{R}_+^{N+1}} \quad t_k R_k(\mathbf{y}_k/t_k, \mathbf{p}_{-k}) \qquad (66)$$
$$\text{subject to} \quad t_k(\Psi_k + \mathbf{1}^T \mathbf{y}_k/t_k) \leq 1$$
$$\mathbf{1}^T \mathbf{y}_k/t_k - P_k \leq 0$$

which is equivalent to (15), since the inequality in the first constraint can be changed to an equality if the denominator of the EE function is affine, and it is convex in $(\mathbf{y}_k, t_k)$, since the perspective of a function preserves convexity [31]. To proceed further, let us define the slack vector $\mathbf{z}_k = \mathbf{y}_k/t_k$ and denote by $\nu_k$ and $\mu_k$ the dual variables associated with the constraints $t_k(\Psi_k + \mathbf{1}^T \mathbf{z}_k) - 1 \leq 0$ and $\mathbf{1}^T \mathbf{z}_k - P_k \leq 0$. Let us also denote $t_k^\star(\mathbf{p}_{-k})$ the optimum value of $t_k$ when the constraint $\mathbf{1}^T \mathbf{z}_k - P_k \leq 0$ on the maximum transmit power is neglected. The KKT conditions of (66) yield

$$\mathbf{z}_k^\star = \mathsf{wf}_k(\mathbf{p}_{-k}, \nu_k^\star + \mu_k^\star) \qquad (67)$$

where the waterfilling operator is defined as in (7). Now, it is worth observing that $R_k(\mathbf{z}_k, \mathbf{p}_{-k})$ is a monotonically increasing function of $z_k(n)$ for any $n$. It follows that, at the optimum point, either the first or the second constraint in (66) are satisfied with equality. Hence, from (67), using the complementary slackness conditions it follows that, if $1/t_k^\star(\mathbf{p}_{-k}) - \Psi_k \leq P_k$, then $\mu_k^\star = 0$, whereas $\nu_k^\star \neq 0$ and such that

$$\mathbf{1}^T \mathsf{wf}_k(\mathbf{p}_{-k}, \nu_k^\star) = \frac{1}{t_k^\star(\mathbf{p}_{-k})} - \Psi_k. \qquad (68)$$

On the other hand, if $1/t_k^\star(\mathbf{p}_{-k}) - \Psi_k > P_k$, then $\nu_k^\star = 0$ whereas $\mu_k^\star \neq 0$ and such that $\mathbf{1}^T \mathsf{wf}_k(\mathbf{p}_{-k}, \mu_k^\star) = P_k$. Consequently, it turns out that $\mathbf{z}_k^\star$ has the following waterfilling structure:

$$\mathbf{z}_k^\star = \mathsf{wf}_k(\mathbf{p}_{-k}, \lambda_k^\star) \qquad (69)$$

where $\lambda_k^\star$ must satisfy the power constraint

$$\mathbf{1}^T \mathbf{z}_k^\star = \min\left\{P_k, \frac{1}{t_k^\star(\mathbf{p}_{-k})} - \Psi_k\right\}. \qquad (70)$$

Recalling that $\mathbf{z}_k^\star = \mathbf{y}_k^\star/t_k^\star$, from (64) and (65) the results (16) and (17) easily follow.

## APPENDIX B
## PROOF OF THEOREM 2

We start observing that $g_k(\mathbf{p}_k, \mathbf{p}_{-k})$ in (28) is convex and continuously differentiable with respect to $\mathbf{p}_k$ for any possible vector $\mathbf{p}_{-k} \in \mathcal{P}_{-k}$. Therefore, we can make use of [22, Theorem 1] and state that $\mathbf{p}^\star$ is a solution of $\mathsf{QVI}(\mathcal{Q}, \mathbf{F})$ if and only if there exist some vectors $\boldsymbol{\nu}^\star \in \mathbb{R}_+^K$ and $\boldsymbol{\mu}^\star \in \mathbb{R}_+^K$ satisfying the following KKT conditions for any $k$ [2]:

$$\mathbf{F}_k\left(\mathbf{p}_k^\star, \mathbf{p}_{-k}^\star\right) + \nu_k^\star \mathbf{1}_N + \mu_k^\star \mathbf{1}_N = \mathbf{0}_N \tag{71}$$

with

$$0 \leq \nu_k^\star \perp g_k\left(\mathbf{p}_k^\star, \mathbf{p}_{-k}^\star\right) \leq 0 \tag{72}$$
$$0 \leq \mu_k^\star \perp h_k\left(\mathbf{p}_k^\star\right) \leq 0. \tag{73}$$

From (71) – (73), using (2) and (28), we can easily verify that $\mathbf{p}^\star$ is thus a solution of $\mathsf{QVI}(\mathcal{Q}, \mathbf{F})$ if and only if

$$\mathbf{p}_k^\star = \mathsf{wf}_k(\mathbf{p}_{-k}^\star, \lambda_k^\star) \tag{74}$$

with $\lambda_k^\star$ being such that

$$\mathbf{1}^T \mathsf{wf}_k(\mathbf{p}_{-k}^\star, \lambda_k^\star) = \min\left\{P_k, \frac{1}{t_k^\star(\mathbf{p}_{-k}^\star)} - \Psi_k\right\}. \tag{75}$$

The above condition coincides exactly with the condition for an NE in $\mathcal{G}_E$ as stated in Proposition 3. This means that $\mathbf{p}^\star$ is also an NE of $\mathcal{G}_E$.

## APPENDIX C
## PROOF OF PROPOSITION 5

Let us consider the following problem:

$$\mathbf{z}_k^\star = \Pi_{\overline{\mathcal{Q}}_k(\mathbf{p}_{-k})}\left(-\boldsymbol{\xi}_k - \sum_{i \neq k} \mathbf{D}_{k,i} \mathbf{p}_i\right) \tag{76}$$

which corresponds to the Euclidean projection of the vector $-\boldsymbol{\xi}_k - \sum_{i \neq k} \mathbf{D}_{k,i} \mathbf{p}_i$ onto the simplex $\overline{\mathcal{Q}}_k(\mathbf{p}_{-k})$ defined as in (33). Mathematically, $\mathbf{z}_k^\star$ is computed looking for the solution of the following minimization problem:

$$\mathbf{z}_k^\star = \arg\min_{\mathbf{z}_k} \ \frac{1}{2}\left\|\mathbf{z}_k - \left(-\boldsymbol{\xi}_k - \sum_{i \neq k} \mathbf{D}_{k,i} \mathbf{p}_i\right)\right\|^2 \tag{77}$$
$$\text{subject to} \quad \mathbf{z}_k \in \overline{\mathcal{Q}}_k(\mathbf{p}_{-k})$$

which is in a convex form for any given $\mathbf{p}_{-k}$. Therefore, the associated Lagrangian is

$$\mathcal{L}(\mathbf{z}_k, \boldsymbol{\nu}_k, \theta_k) = \frac{1}{2}\sum_{n=1}^N (z_k(n) + \zeta_k(n))^2$$
$$- \theta_k \left(\sum_{n=1}^N z_k(n) - \min\left\{P_k, \frac{1}{t_k^\star(\mathbf{p}_{-k})} - \Psi_k\right\}\right)$$
$$- \sum_{n=1}^N \nu_k(n) z_k(n) \tag{78}$$

---
[2] Observe that, according to [22], the gradient of the constraint functions, $g_k(\cdot,\cdot)$ for any $k$ in (71), are computed only with respect to the first argument $\mathbf{p}_k$, while the second argument is kept as a constant. Hence, we have $\nabla_{\mathbf{p}_k} g_k(\mathbf{p}_k, \mathbf{p}_{-k}) = \mathbf{1}_N$.

with $\zeta_k(n) = \xi_k(n) + \sum_{i\neq k} D_{k,i}(n) p_i(n)$, and the problem solution $\mathbf{z}_k^\star$ is such that the following KKT conditions are satisfied for any $n$:

$$z_k^\star(n) + \zeta_k(n) - \theta_k^\star - \nu_k^\star(n) = 0 \tag{79}$$
$$z_k^\star(n) \geq 0 \qquad \nu_k^\star(n) \geq 0 \qquad \nu_k^\star(n) z_k^\star(n) = 0 \tag{80}$$
$$\sum_{n=1}^N z_k^\star(n) = \min\left\{P_k, \frac{1}{t_k^\star(\mathbf{p}_{-k})} - \Psi_k\right\}. \tag{81}$$

From (79) and (80), if $z_k^\star(n) > 0$, we obtain

$$z_k^\star(n) = \theta_k^\star - \zeta_k(n) > 0 \tag{82}$$

with $\zeta_k(n) > 0$. Thus, $\theta_k^\star > 0$ whenever $z_k^\star(n) > 0$. On the other hand, if $z_k^\star(n) = 0$, we get

$$-\nu_k^\star(n) = \theta_k^\star - \zeta_k(n) \leq 0. \tag{83}$$

Finally, from (81) – (83) we obtain

$$z_k^\star(n) = [\theta_k^\star - \zeta_k(n)]^+ \tag{84}$$

with $\theta_k^\star$ being such that

$$\mathbf{1}^T \mathbf{z}_k^\star = \min\left\{P_k, \frac{1}{t_k^\star(\mathbf{p}_{-k})} - \Psi_k\right\}. \tag{85}$$

Setting $\theta_k^\star = 1/\lambda_k^\star$, it follows that (84) and (85) are equivalent to (69) and (70), respectively. This means that the parameter-free optimization problem in (66) is equivalent to the minimization problem in (77). Since the optimal power allocation profile $\mathbf{p}_k^\star$ of player $k$ is always an instance of $\mathbf{z}_k^\star$, it follows that $\mathbf{p}^\star$ is a NE of the energy-efficient maximization problem if and only if it is such that:

$$\mathbf{p}^\star = \Pi_{\overline{\mathcal{Q}}(\mathbf{p}^\star)}\left(-\boldsymbol{\xi}_k - \sum_{i \neq k} \mathbf{D}_{k,i} \mathbf{p}_i\right). \tag{86}$$

The claim of Proposition 5 is thus proved.

## APPENDIX D
## PROOF OF THEOREM 3

The uniqueness result of Theorem 3 is a consequence of the following theorem [28, Theorem 4.1] (see also [32, Theorem 9]).

**Theorem 4.** *Let the following assumptions hold.*
- *The operator $\mathbf{F}$ is strongly monotone $\forall \mathbf{p}, \mathbf{p}' \in \mathcal{P}$, i.e.,*

$$(\mathbf{p} - \mathbf{p}')^T (\mathbf{F}(\mathbf{p}) - \mathbf{F}(\mathbf{p}')) \geq \beta \|\mathbf{p} - \mathbf{p}'\|^2 \tag{87}$$

 *with $\beta > 0$ being the strong monotonicity constant;*
- *The operator $\mathbf{F}$ is Lipschitz continuous $\forall \mathbf{p}, \mathbf{p}' \in \mathcal{P}$ with modulus $L > 0$, i.e.,*

$$\|\mathbf{F}(\mathbf{p}) - \mathbf{F}(\mathbf{p}')\| \leq L \|\mathbf{p} - \mathbf{p}'\|; \tag{88}$$

- *There exists a constant $\delta \geq 0$ such that $\forall \mathbf{z}, \mathbf{p}, \mathbf{p}' \in \mathcal{P}$*

$$\|\Pi_{\mathcal{S}(\mathbf{p})}(\mathbf{z}) - \Pi_{\mathcal{S}(\mathbf{p}')}(\mathbf{z})\| \leq \delta \|\mathbf{p} - \mathbf{p}'\|$$

 *with $\delta < \beta/L$.*

*Then, $\mathsf{QVI}(\mathcal{S}, \mathbf{F})$ has a unique solution.*

The first condition easily follows from [14, Proposition 2] in which it is stated that if $\mathbf{B}$ in (40) is positive definite then the operator $\mathbf{F}$ is strongly monotone $\forall \mathbf{p}, \mathbf{p}' \in \mathcal{P}$ with

$$\beta = \frac{\lambda_{\min}(\widetilde{\mathbf{B}})}{\max_k \max_n \{(\tilde{\varsigma}_k(n))^2\}} \tag{89}$$

where $\widetilde{\mathbf{B}}$ represent the symmetric part of $\mathbf{B}$ and

$$\tilde{\varsigma}_k(n) = \frac{\sigma_k^2(n)}{|H_{k,k}(n)|^2} + \sum_i \frac{|H_{k,i}(n)|^2}{|H_{k,k}(n)|^2} P_i. \tag{90}$$

To prove that $\mathbf{F}$ is Lipschitz continuous, we make use of (10) to get

$$\|\mathbf{F}_k(\mathbf{p}) - \mathbf{F}_k(\mathbf{p}')\| \leq \\ \left\| \left\{ -\frac{\sum_{i=1}^K D_{k,i}(n) p_i(n) - \sum_{i=1}^K D_{k,i}(n) p_i'(n)}{\xi_k^2(n)} \right\}_{n=1}^N \right\| \tag{91}$$

since $\xi_k(n) + \sum_{i=1}^K D_{k,i}(n) p_i(n) \geq \xi_k(n)$. Letting $\widetilde{\mathbf{A}}_{k,i}$ be a diagonal matrix with elements

$$\left[ \widetilde{\mathbf{A}}_{k,i} \right]_{n,n} = \frac{D_{k,i}(n)}{\xi_k^2(n)} = \frac{|H_{k,i}(n)|^2 |H_{k,k}(n)|^2}{\sigma_k^4(n)} \tag{92}$$

from (91) we obtain

$$\|\mathbf{F}_k(\mathbf{p}) - \mathbf{F}_k(\mathbf{p}')\| \leq \left\| \sum_{i=1}^K \widetilde{\mathbf{A}}_{k,i}(\mathbf{p}_i - \mathbf{p}_i') \right\|. \tag{93}$$

Observe now that

$$\left\| \sum_{i=1}^K \widetilde{\mathbf{A}}_{k,i}(\mathbf{p}_i - \mathbf{p}_i') \right\| \leq \sum_{i=1}^K \max_n \left\{ \left[ \widetilde{\mathbf{A}}_{k,i} \right]_{n,n} \right\} \|\mathbf{p}_i - \mathbf{p}_i'\| \tag{94}$$

so that from (93) one gets

$$\|\mathbf{F}_k(\mathbf{p}) - \mathbf{F}_k(\mathbf{p}')\| \leq \sum_{i=1}^K [\mathbf{A}]_{k,i} \|\mathbf{p}_i - \mathbf{p}_i'\| \tag{95}$$

with $\mathbf{A}$ being defined as in (39). Then, we may write

$$\|\mathbf{F}(\mathbf{p}) - \mathbf{F}(\mathbf{p}')\|^2 \leq \sum_{k=1}^K \left( \sum_{i=1}^K [\mathbf{A}]_{k,i} \|\mathbf{p}_i - \mathbf{p}_i'\| \right)^2 \tag{96}$$

or, equivalently,

$$\|\mathbf{F}(\mathbf{p}) - \mathbf{F}(\mathbf{p}')\|^2 \leq \|\mathbf{A}\mathbf{x}\|^2 \tag{97}$$

where

$$\mathbf{x} = [\|\mathbf{p}_1 - \mathbf{p}_1'\|, \cdots, \|\mathbf{p}_K - \mathbf{p}_K'\|]^T. \tag{98}$$

Observe that

$$\|\mathbf{A}\mathbf{x}\| \leq \|\mathbf{A}\| \|\mathbf{x}\| = \|\mathbf{A}\| \|\mathbf{p} - \mathbf{p}'\| \tag{99}$$

with $\|\mathbf{A}\| = \max_{\|\mathbf{x}\|=1} \{\|\mathbf{A}\mathbf{x}\|\} = \sqrt{\lambda_{\max}(\mathbf{A}^H \mathbf{A})}$ being the induced norm of matrix $\mathbf{A}^H \mathbf{A}$. Therefore, we may write

$$\|\mathbf{F}(\mathbf{p}) - \mathbf{F}(\mathbf{p}')\| \leq \sqrt{\lambda_{\max}(\mathbf{A}^H \mathbf{A})} \|\mathbf{p} - \mathbf{p}'\| \tag{100}$$

which proves that $\mathbf{F}$ is Lipschitz continuous $\forall \mathbf{p}, \mathbf{p}' \in \mathcal{P}$ with modulus $L = \sqrt{\lambda_{\max}(\mathbf{A}^H \mathbf{A})}$.

We now proceed proving that the second condition of Theorem 3 holds true if (42) is satisfied. To this end, we start observing that

$$\left\| \Pi_{\mathcal{S}(\mathbf{p})}(\mathbf{z}) - \Pi_{\mathcal{S}(\mathbf{p}')}(\mathbf{z}) \right\|^2 = \\ \sum_{k \in \mathcal{K}_E} \left\| \Pi_{\mathcal{S}_k(\mathbf{p}_{-k})}(\mathbf{z}_k) - \Pi_{\mathcal{S}_k(\mathbf{p}'_{-k})}(\mathbf{z}_k) \right\|^2. \tag{101}$$

Since the upper boundaries of $\mathcal{S}_k(\mathbf{p}_{-k})$ and $\mathcal{S}_k(\mathbf{p}'_{-k})$, denoted with $\overline{\mathcal{S}}_k(\mathbf{p}_{-k})$ and $\overline{\mathcal{S}}_k(\mathbf{p}'_{-k})$ respectively, are two parallel hyperplanes in $\mathbf{p}_k$, i.e. either $\mathcal{S}_k(\mathbf{p}_{-k}) \subset \mathcal{S}_k(\mathbf{p}'_{-k})$ or $\mathcal{S}_k(\mathbf{p}'_{-k}) \subseteq \mathcal{S}_k(\mathbf{p}_{-k})$, we may write

$$\sum_{k \in \mathcal{K}_E} \left\| \Pi_{\mathcal{S}_k(\mathbf{p}_{-k})}(\mathbf{z}_k) - \Pi_{\mathcal{S}_k(\mathbf{p}'_{-k})}(\mathbf{z}_k) \right\|^2 \leq \\ \sum_{k \in \mathcal{K}_E} \left\| \Pi_{\overline{\mathcal{S}}_k(\mathbf{p}_{-k})}(\mathbf{z}_k) - \Pi_{\overline{\mathcal{S}}_k(\mathbf{p}'_{-k})}(\mathbf{z}_k) \right\|^2. \tag{102}$$

Observing that

$$\sum_{k \in \mathcal{K}_E} \left\| \Pi_{\overline{\mathcal{S}}_k(\mathbf{p}_{-k})}(\mathbf{z}_k) - \Pi_{\overline{\mathcal{S}}_k(\mathbf{p}'_{-k})}(\mathbf{z}_k) \right\|^2 \leq \\ \leq \sum_{k \in \mathcal{K}_E} \left( \frac{1}{t_k(\mathbf{p}'_{-k})} - \frac{1}{t_k(\mathbf{p}_{-k})} \right)^2 = \|\mathbf{\Omega}(\mathbf{p}) - \mathbf{\Omega}(\mathbf{p}')\|^2$$

and using (101) and (102) yields

$$\left\| \Pi_{\mathcal{S}(\mathbf{p})}(\mathbf{z}) - \Pi_{\mathcal{S}(\mathbf{p}')}(\mathbf{z}) \right\|^2 \leq \|\mathbf{\Omega}(\mathbf{p}) - \mathbf{\Omega}(\mathbf{p}')\|^2$$

from which taking into account (42) we eventually obtain

$$\left\| \Pi_{\mathcal{S}(\mathbf{p})}(\mathbf{z}) - \Pi_{\mathcal{S}(\mathbf{p}')}(\mathbf{z}) \right\| \leq \delta \|\mathbf{p} - \mathbf{p}'\| \tag{103}$$

as required by the second condition of Theorem 3.

## APPENDIX E
## PROOF OF PROPOSITION 8

The proof of Proposition 8 follows from the more general result provided by [23, Theorem 3], according to which the SPA leads to the solution of a generic $\mathrm{QVI}(\mathcal{S}, \mathbf{F})$ with $\mathcal{S}(\mathbf{p}) = \prod_{k=1}^K \mathcal{S}_k(\mathbf{p}_{-k})$, and $\mathcal{S}_k(\mathbf{p}_{-k})$ defined as in (35), if the following conditions are satisfied:

- $\mathbf{F}$ is continuous in $\mathbf{p}$ and $h_k(\mathbf{p}_k)$ is continuously differentiable and convex in $\mathbf{p}_k$;
- $g_k(\mathbf{p}_k, \mathbf{p}_{-k})$ is continuously differentiable and convex in $\mathbf{p}_k$;
- $g_k(\mathbf{p}_k, \mathbf{p}_{-k})$ is continuous in $\mathbf{p}$.

From (2) and (10), it follows that the first condition is verified for the problem at hand. The second condition is also met since $g_k(\mathbf{p}_k, \mathbf{p}_{-k})$ in (28) is an affine function of $\mathbf{p}_k$ for any given $\mathbf{p}_{-k}$. Therefore, we are only left with proving that $g_k(\mathbf{p}_k, \mathbf{p}_{-k})$ is continuous in $\mathbf{p}$. From (28), this amounts to showing that $t_k^\star(\mathbf{p}_{-k})$ is continuous $\forall \mathbf{p}_{-k} \in \mathcal{P}_{-k}$ with $\mathcal{P}_{-k} = \prod_{i \neq k} \mathcal{P}_i$. This is proved by contradiction as follows. Assume that there exists a pair of arbitrary sequences $\{\mathbf{z}_1^{(j)}\}$, $\{\mathbf{z}_2^{(j)}\}$ with $\mathbf{z}_1^{(j)}, \mathbf{z}_2^{(j)} \in \mathbb{R}^{N(K-1)}$, $\forall j$ such that

$$\lim_{j \to \infty} \mathbf{z}_1^{(j)} = \mathbf{p}_{-k}$$

and
$$\lim_{j\to\infty} \mathbf{z}_2^{(j)} = \mathbf{p}_{-k}$$

with
$$\lim_{j\to\infty} t_k^\star(\mathbf{z}_1^{(j)}) \neq \lim_{j\to\infty} t_k^\star(\mathbf{z}_2^{(j)}). \quad (104)$$

From (18) – (19), this implies that there exist two distinct values of the Lagrangian multiplier $\nu_k^\star$, namely, $\nu_{k,1}^\star$ and $\nu_{k,2}^\star$, such that the condition (20) is fulfilled. At the same time, from (20) it follows that any instance of $\nu_k^\star$ is such that

$$\nu_k^\star = \frac{R_k\left(\mathbf{z}_k(\nu_k^\star), \mathbf{p}_{-k}\right)}{\Psi_k + \mathbf{1}^T \mathbf{z}_k(\nu_k^\star)} \quad (105)$$

which is nothing else than the maximum value of the EE function $E_k(\mathbf{p}_k, \mathbf{p}_{-k}) = t_k R_k(\mathbf{p}_k, \mathbf{p}_{-k})$ in (23). Since $E_k(\mathbf{p}_k, \mathbf{p}_{-k})$ is strictly quasiconcave [19], it follows that $\nu_k^\star$ must be unique. Accordingly, we must conclude that there are no distinct sequences $\{\mathbf{z}_1^{(j)}\}$ and $\{\mathbf{z}_2^{(j)}\}$ such that (104) is satisfied. This concludes the proof.

## APPENDIX F
## PROOF OF PROPOSITION 9

We start observing that $\mathbf{p}^\star$ is a solution of $\mathsf{QVI}(\mathcal{S}, \mathbf{F})$ if and only if there exist some vectors $\boldsymbol{\nu}^\star \in \mathbb{R}_+^K$ and $\boldsymbol{\mu}^\star \in \mathbb{R}_+^K$ such that the following KKT conditions are satisfied [22], [23]:

$$\mathbf{F}_k(\mathbf{p}^\star) + \nu_k^\star \mathbf{1}_N + \mu_k^\star \mathbf{1}_N = \mathbf{0}_N \quad (106)$$
$$0 \leq \mu_k^\star \perp h_k(\mathbf{p}_k^\star) \leq 0 \quad (107)$$

with $\nu_k^\star = 0$ if $k \in \mathcal{K}_R$, and

$$0 \leq \nu_k^\star \perp g_k(\mathbf{p}_k^\star, \mathbf{p}_{-k}^\star) \leq 0 \quad (108)$$

if $k \in \mathcal{K}_E$. Next, we look for a procedure that allows to compute a triplet $(\mathbf{p}^\star, \boldsymbol{\nu}^\star, \boldsymbol{\mu}^\star)$ in a distributed manner. To this end, we observe that the solution $\mathbf{p}^{\mathsf{VI}}(\boldsymbol{\gamma})$ of (56) must be such that there exists some vector $\boldsymbol{\chi}^\star \in \mathbb{R}_+^K$ satisfying

$$\mathbf{F}_k(\mathbf{p}^{\mathsf{VI}}(\boldsymbol{\gamma})) + \gamma_k \mathbf{1}_N + \chi_k^\star \mathbf{1}_N = \mathbf{0}_N \quad (109)$$
$$0 \leq \chi_k^\star \perp h_k(\mathbf{p}_k^{\mathsf{VI}}(\boldsymbol{\gamma})) \leq 0. \quad (110)$$

Let $\boldsymbol{\gamma}^\star$ be the solution of the following nonlinear complementarity problem $\mathsf{NCP}(\boldsymbol{\Phi})$, with $\boldsymbol{\Phi}$ defined as in (57):

$$\begin{aligned}
\text{find} \quad & \boldsymbol{\gamma} \succeq \mathbf{0} \\
\text{subject to} \quad & \boldsymbol{\Phi}(\boldsymbol{\gamma}) \succeq \mathbf{0} \\
& 0 \leq \boldsymbol{\gamma} \perp \boldsymbol{\Phi}(\boldsymbol{\gamma}) \geq \mathbf{0}.
\end{aligned} \quad (111)$$

We can easily see that the triplet $(\mathbf{p}^{\mathsf{VI}}(\boldsymbol{\gamma}^\star), \boldsymbol{\gamma}^\star, \boldsymbol{\chi}^\star)$ satisfies the KKT conditions of $\mathsf{QVI}(\mathcal{S}, \mathbf{F})$ (see also [29] for more details on finite-dimensional VIs and complementarity problems).

## APPENDIX G

According to Proposition 10, the main condition for the convergence of the proposed algorithm is the co-coercivity of the operator $\boldsymbol{\Phi}$, i.e., there exists a constant $\kappa$ such that

$$(\boldsymbol{\gamma}_1 - \boldsymbol{\gamma}_2)^T (\boldsymbol{\Phi}(\boldsymbol{\gamma}_1) - \boldsymbol{\Phi}(\boldsymbol{\gamma}_2)) \geq \kappa \|\boldsymbol{\Phi}(\boldsymbol{\gamma}_1) - \boldsymbol{\Phi}(\boldsymbol{\gamma}_2)\|^2 \quad (112)$$

where, from (38) and (57), we can observe that

$$(\boldsymbol{\gamma}_1 - \boldsymbol{\gamma}_2)^T (\boldsymbol{\Phi}(\boldsymbol{\gamma}_1) - \boldsymbol{\Phi}(\boldsymbol{\gamma}_2)) =$$
$$= (\boldsymbol{\gamma}_1 - \boldsymbol{\gamma}_2)^T (\boldsymbol{\Omega}(\mathbf{p}^{\mathsf{VI}}(\boldsymbol{\gamma}_1)) - \boldsymbol{\Omega}(\mathbf{p}^{\mathsf{VI}}(\boldsymbol{\gamma}_2)))$$
$$- \sum_{k=1}^{K} (\gamma_{1,k} - \gamma_{2,k})(\mathbf{1}^T \mathbf{p}^{\mathsf{VI}}(\boldsymbol{\gamma}_1) - \mathbf{1}^T \mathbf{p}^{\mathsf{VI}}(\boldsymbol{\gamma}_2)). \quad (113)$$

Let us assume that the uniqueness conditions presented in Theorem 3 are fulfilled and consider the right side of (112). Exploiting the Cauchy-Schwarz inequality, we may write

$$\|\boldsymbol{\Phi}(\boldsymbol{\gamma}_1) - \boldsymbol{\Phi}(\boldsymbol{\gamma}_2)\|^2 \leq$$
$$\|\boldsymbol{\Omega}(\mathbf{p}(\boldsymbol{\gamma}_1)) - \boldsymbol{\Omega}(\mathbf{p}(\boldsymbol{\gamma}_2))\|^2 + \|\mathbf{p}^{\mathsf{VI}}(\boldsymbol{\gamma}_1) - \mathbf{p}^{\mathsf{VI}}(\boldsymbol{\gamma}_2)\|^2 \quad (114)$$

where, from (42), one gets

$$\|\boldsymbol{\Phi}(\boldsymbol{\gamma}_1) - \boldsymbol{\Phi}(\boldsymbol{\gamma}_2)\|^2 \leq$$
$$\delta^2 \|\mathbf{p}^{\mathsf{VI}}(\boldsymbol{\gamma}_1) - \mathbf{p}^{\mathsf{VI}}(\boldsymbol{\gamma}_2)\|^2 + \|\mathbf{p}^{\mathsf{VI}}(\boldsymbol{\gamma}_1) - \mathbf{p}^{\mathsf{VI}}(\boldsymbol{\gamma}_2)\|^2 =$$
$$(1 + \Gamma^{-2}) \|\mathbf{p}^{\mathsf{VI}}(\boldsymbol{\gamma}_1) - \mathbf{p}^{\mathsf{VI}}(\boldsymbol{\gamma}_2)\|^2 \quad (115)$$

with $\Gamma = L/\beta$. To proceed further, we take advantage of the results in [14, Proposition 8] wherein it is proved that

$$-\sum_{k=1}^{K} (\gamma_{1,k} - \gamma_{2,k})(\mathbf{1}^T \mathbf{p}^{\mathsf{VI}}(\boldsymbol{\gamma}_1) - \mathbf{1}^T \mathbf{p}^{\mathsf{VI}}(\boldsymbol{\gamma}_2)) \geq$$
$$\beta \|\mathbf{p}^{\mathsf{VI}}(\boldsymbol{\gamma}_1) - \mathbf{p}^{\mathsf{VI}}(\boldsymbol{\gamma}_2)\|^2 \geq \frac{\beta}{1 + \Gamma^{-2}} \|\boldsymbol{\Phi}(\boldsymbol{\gamma}_1) - \boldsymbol{\Phi}(\boldsymbol{\gamma}_2)\|^2. \quad (116)$$

At this point, we are only left with the scalar product:

$$(\boldsymbol{\gamma}_1 - \boldsymbol{\gamma}_2)^T \left[\boldsymbol{\Omega}(\mathbf{p}^{\mathsf{VI}}(\boldsymbol{\gamma}_1)) - \boldsymbol{\Omega}(\mathbf{p}^{\mathsf{VI}}(\boldsymbol{\gamma}_2))\right] =$$
$$\sum_{k=1}^{K} (\gamma_{1,k} - \gamma_{2,k}) \left(\frac{1}{t_k^\star(\mathbf{p}_{-k}^{\mathsf{VI}}(\boldsymbol{\gamma}_1))} - \frac{1}{t_k^\star(\mathbf{p}_{-k}^{\mathsf{VI}}(\boldsymbol{\gamma}_2))}\right). \quad (117)$$

From (18), one gets

$$[\boldsymbol{\Omega}(\mathbf{p}^{\mathsf{VI}}(\boldsymbol{\gamma}))]_k = \frac{1}{t_k^\star(\mathbf{p}_{-k}^{\mathsf{VI}}(\boldsymbol{\gamma}))} - \Psi_k = \mathbf{1}^T \mathsf{wf}_k(\mathbf{p}_{-k}^{\mathsf{VI}}(\boldsymbol{\gamma}), \nu_k^\star)$$

which represents the radiated power we would have at the transmitter $k$ when the EE is maximized given $\mathbf{p}_{-k}^{\mathsf{VI}}(\boldsymbol{\gamma})$. The lack of an explicit form for $t_k^\star(\mathbf{p}_{-k}^{\mathsf{VI}}(\boldsymbol{\gamma}))$ makes it hard to study in a rigorous way the co-coercivity of $(\boldsymbol{\Omega}(\mathbf{p}^{\mathsf{VI}}(\boldsymbol{\gamma}))$ with respect to $\boldsymbol{\gamma}$. To partially fulfill this lack, the following heuristic line of reasoning is used to get some insights. Intuitively speaking, when the penalty coefficients in $\boldsymbol{\gamma}$ increase, the interfering powers in $\mathbf{p}_{-k}^{\mathsf{VI}}(\boldsymbol{\gamma})$ decrease and the user $k$ experiences larger SINRs. According to the waterfilling principle, a larger SINR implies a larger radiated power, or, in other words,

$$(\boldsymbol{\gamma}_1 - \boldsymbol{\gamma}_2)^T (\boldsymbol{\Omega}(\mathbf{p}^{\mathsf{VI}}(\boldsymbol{\gamma}_1)) - \boldsymbol{\Omega}(\mathbf{p}^{\mathsf{VI}}(\boldsymbol{\gamma}_2))) \geq 0. \quad (118)$$

Then, collecting (113) and (115)-(116), one gets

$$(\boldsymbol{\gamma}_1 - \boldsymbol{\gamma}_2)^T (\boldsymbol{\Phi}(\boldsymbol{\gamma}_1) - \boldsymbol{\Phi}(\boldsymbol{\gamma}_2)) \geq \frac{\beta}{1 + \Gamma^{-2}} \|\boldsymbol{\Phi}(\boldsymbol{\gamma}_1) - \boldsymbol{\Phi}(\boldsymbol{\gamma}_2)\|^2. \quad (119)$$

## References

[1] Cisco, "Cisco visual networking index: Global mobile data traffic forecast update, 2010 - 2015," *Whitepaper*, Feb. 2011.

[2] J. Hoydis, M. Kobayashi, and M. Debbah, "Green small-cell networks," *IEEE Veh. Technol. Mag.*, vol. 6, no. 1, pp. 37–43, Mar. 2011.

[3] S. Lasaulce and H. Tembine, *Game Theory and Learning for Wireless Networks: Fundamentals and Applications*. Waltham, MA, USA: Academic Press, 2011.

[4] W. Yu, G. Ginis, and J. M. Cioffi, "Distributed multiuser power control for digital subscriber lines," *IEEE J. Sel. Areas Commun.*, vol. 20, no. 5, pp. 1105–1115, Aug. 2002.

[5] N. Yamashita and Z.-Q. Luo, "A nonlinear complementarity approach to multiuser power control for digital subscriber lines," *Optim. Methods and Software*, vol. 19, no. 5, pp. 633–652, 2004.

[6] R. Cendrillon, J. Huang, M. Chiang, and M. Moonen, "Autonomous spectrum balancing for digital subscriber lines," *IEEE Trans. Signal Process.*, vol. 55, no. 8, pp. 4241–4257, Aug. 2007.

[7] G. Scutari, D. P. Palomar, and S. Barbarossa, "Competitive design of multiuser MIMO systems based on game theory: A unified view," *IEEE J. Sel. Areas Commun.*, vol. 26, no. 7, pp. 1089–1103, Sep. 2008.

[8] Y. Shi, J. Wang, K. Letaief, and R. Mallik, "A game-theoretic approach for distributed power control in interference relay channels," *IEEE Trans. Wireless Commun.*, vol. 8, no. 6, pp. 3151–3161, Jun. 2009.

[9] S. Ren and M. van der Schaar, "Distributed power allocation in multi-user multi-channel cellular relay networks," *IEEE Trans. Wireless Commun.*, vol. 9, no. 6, pp. 1952–1964, Jun. 2010.

[10] D. T. Ngo, L. B. Le, T. Le-Ngoc, E. Hossain, and D. I. Kim, "Distributed interference management in two-tier CDMA femtocell networks," *IEEE Trans. Wireless Commun.*, vol. 11, no. 3, pp. 979–989, Mar. 2012.

[11] G. Scutari, D. P. Palomar, and S. Barbarossa, "Optimal linear precoding strategies for wideband noncooperative systems based on game theory — part I: Nash equilibria," *IEEE Trans. Signal Process.*, vol. 56, no. 3, pp. 1230–1249, Mar. 2008.

[12] ——, "Optimal linear precoding strategies for wideband noncooperative systems based on game theory — part II: Algorithms," *IEEE Trans. Signal Process.*, vol. 56, no. 3, pp. 1250–1267, Mar. 2008.

[13] G. Scutari, D. P. Palomar, F. Facchinei, and J.-S. Pang, "Convex optimization, game theory, and variational inequality theory," *IEEE Signal Process. Mag.*, vol. 27, no. 3, pp. 35–49, May 2010.

[14] J.-S. Pang, G. Scutari, D. P. Palomar, and F. Facchinei, "Design of cognitive radio systems under temperature-interference constraints: A variational inequality approach," *IEEE Trans. Signal Process.*, vol. 58, no. 6, pp. 3251–3271, Jun. 2010.

[15] S. Verdú, "On channel capacity per unit cost," *IEEE Trans. Inf. Theory*, vol. 36, no. 5, pp. 1019–1030, Sep. 1990.

[16] E.-V. Belmega and S. Lasaulce, "Energy-efficient precoding for multiple-antenna terminals," *IEEE Trans. Signal Process.*, vol. 59, no. 1, pp. 329–340, Jan. 2011.

[17] C. Isheden, Z. Chong, E. Jorswieck, and G. Fettweis, "Framework for link-level energy efficiency optimization with informed transmitter," *IEEE Trans. Wireless Commun.*, vol. 11, no. 8, pp. 2946–2957, Aug. 2012.

[18] R. Xie, F. R. Yu, H. Ji, and Y. Li, "Energy-efficient resource allocation for heterogeneous cognitive radio networks with femtocells," *IEEE Trans. Wireless Commun.*, vol. 11, no. 11, pp. 3910–3920, Nov. 2012.

[19] G. Miao, N. Himayat, G. Li, and S. Talwar, "Distributed interference-aware energy-efficient power optimization," *IEEE Trans. Wireless Commun.*, vol. 10, no. 4, pp. 1323–1333, Apr. 2011.

[20] G. Bacci, E. Belmega, P. Mertikopoulos, and L. Sanguinetti, "Energy-aware competitive power allocation in heterogeneous networks with QoS constraints," *IEEE Trans. Wireless Commun.*, 2015, to appear. [Online]. Available: http://arxiv.org/abs/1408.7004

[21] A. Bensoussan, "Points de Nash dans le cas de fonctionnelles quadratiques et jeux differentiels linéaires a *N* personnes," *SIAM Journal on Control*, vol. 12, no. 3, pp. 460–499, Mar.-Apr. 1974.

[22] F. Facchinei, C. Kanzow, and S. Sagratella, "Solving quasi-variational inequalities via their KKT conditions," *Mathematical Programming*, vol. 144, no. 1-2, pp. 369–412, Feb. 2013.

[23] J.-S. Pang and M. Fukushima, "Quasi-variational inequalities, generalized Nash equilibria, and multi-leader-follower games," *Computational Management Science*, vol. 2, no. 1, pp. 21–56, Jan. 2005.

[24] R. Jagannathan, "Duality for nonlinear fractional programs," *Zeitschrift für Operations Research*, vol. 17, no. 1, pp. 1–3, Feb. 1973.

[25] W. Dinkelbach, "On nonlinear fractional programming," *Management Science*, vol. 13, no. 7, pp. 492–498, Mar. 1967.

[26] K. J. Devlin, *Fundamentals of Contemporary Set Theory*. New York: Springer-Verlag, 1979.

[27] D. Fudenberg and J. Tirole, *Game Theory*. Cambridge, MA: MIT Press, 1991.

[28] Y. Nesterov and L. Scrimali, "Solving strongly monotone variational and quasi-variational inequalities," *Discrete and Continuous Dynamical Systems*, vol. 31, no. 4, pp. 1383–1396, Dec. 2011.

[29] F. Facchinei and J.-S. Pang, *Finite-dimensional variational inequalities and complementarity problems*. New York: Springer, 2003.

[30] P. T. Harker, "Generalized Nash games and quasi-variational inequalities," *European Journal of Operational Research*, vol. 54, no. 1, pp. 81–94, Sep. 1991.

[31] S. Boyd and L. Vandenberghe, *Convex Optimization*. Cambridge, UK: Cambridge University Press, 2004.

[32] M. A. Noor and W. Oettli, "On general nonlinear complementarity problems and quasi-equilibria," *Le Matematiche*, vol. 49, no. 2, pp. 313–331, 1995.